\documentclass[aps,prc,preprint]{revtex4}
\usepackage[]{youngtab}
\usepackage{dsfont}
\usepackage{epsfig}
\usepackage{color}

\newcommand{\nn}{\nonumber}

\begin{document}
\title{Three-nucleon forces in the  $1/N_c$ expansion}

\author{Daniel R. Phillips$^{1}$}
\author{Carlos Schat$^{1,2}$}
\affiliation{$^1$ Institute of Nuclear and Particle Physics and 
Department of Physics and Astronomy, Ohio University, Athens, Ohio 45701, USA;\\
$^2$ CONICET - Departamento de F\'{\i}sica, FCEyN, Universidad de Buenos Aires, 
Ciudad Universitaria, Pab.~1, (1428) Buenos Aires, Argentina.}

\begin{abstract} \vspace*{18pt}
The operator structures that can contribute to three-nucleon forces are
classified  in the $1/N_c$ expansion. At leading order in $1/N_c$ a spin-flavor
independent term is present, as are the spin-flavor structures associated with the
Fujita-Miyazawa three-nucleon force.  Modern phenomenological three-nucleon forces are
thus consistent with this ${\cal O}(N_c)$ leading force, corrections to which are
suppressed by a power series in $1/N_c^2$. A complete basis of operators for the three-nucleon force, 
including all independent momentum structures, is given explicitly up to next-to-leading
order in the $1/N_c$ expansion.
\end{abstract}

\date{\today}

\maketitle

\section{Introduction}

Over the last fifteen years advances in few-body methods and the steady increase in
computational power have enabled numerically accurate calculations of few-nucleon
scattering observables and the spectra of light nuclei.  In the three-nucleon system such
calculations show clear evidence for three-nucleon forces (3NFs) when compared with
experimental data~\cite{KalantarNayestanaki:2011wz,Hammer:2012id}.  The simplest and best
known example of this is that the triton binding energy is underestimated by about 800 keV
if a Hamiltonian with two-nucleon potentials alone is employed~\cite{Fr93}. A similar
underbinding occurs for other light nuclei as
well~\cite{Pudliner:1997ck,Wiringa:2000gb,Pieper:2002ne}.  (Although, see
Ref.~\cite{Nogga:2004ab} for a study of the dependence of this conclusion on the
resolution scale at which the NN potential is defined.) Indeed, the role of three-nucleon
forces in the spectra of light nuclei has been a subject of intense investigation during
this period (see, e.g.~\cite{Pieper:2001mp,Na07,Ma12}, as well as
Ref.~\cite{Hammer:2012id}). Recently, state-of-the-art treatments of the role of 3NFs in
heavier nuclei show that they could play a role in determining the location of the
neutron-drip line in the oxygen and calcium isotopes~\cite{Otsuka:2009cs,Holt:2010yb}, and
in extending the half life of Carbon-14~\cite{Holt:2009uk}.  

Historically, 3NFs were first derived in the classic paper of Fujita and
Miyazawa~\cite{FM57}.  There, a 3NF due to the exchange of two pions was computed. This
3NF still forms a key portion of the 3NFs employed today, appearing, for example, in the
Urbana three-nucleon force~\cite{Carlson:1983kq,CS98}:
\begin{equation}
V_{ijk}=V_{ijk}^{2 \pi} + V_{ijk}^R \ ,
\label{eq:Urbana}
\end{equation}
with~\cite{Hu99}
\begin{eqnarray}
&& V_{ijk}^{2 \pi}=\tilde{A}_{2 \pi} \frac{\sigma_i \cdot {\bf k}_1 \sigma_k \cdot {\bf k}_2}{({\bf k}_1^2 + m_\pi^2)({\bf k}_2^2 + m_\pi^2)}\nonumber\\
&& \qquad \left[(a +  b \, {\bf k}_1 \cdot {\bf k}_2) \tau_i \cdot \tau_k + d \, \tau_i \cdot (\tau_j \times \tau_k) \sigma_j \cdot ({\bf k}_1 \times {\bf k}_2 )\right] \ .
\label{eq:FM}
\end{eqnarray}
Here ${\bf k}_{1,2}$ are the momenta of the two pions in the exchange, $\sigma$ and $\tau$
are the usual Pauli matrices for nucleon spin and isospin, and the coefficients $a$, $b$, and
$d$ represent the strength of s-wave and p-wave $\pi$N scattering. If, as was assumed by
Fujita and Miyazawa, we take the p-wave pieces to arise from the spin-3/2, isospin-3/2,
$\pi$N channel, where the $\Delta(1232)$ resides, we have $b=4d$.

Meanwhile, the term $V_{ijk}^R$ in Eq.~(\ref{eq:Urbana}) is spin and isospin independent,
and produces repulsion. The strength of this term, and the overall strength of $V_{ijk}^{2
\pi}$, are adjusted so that calculations with the AV18 NN potential and this 3NF reproduce
the triton binding energy and ``\ldots provide additional repulsion in hypernetted-chain
variational calculations of nuclear matter near equilibrium density"~\cite{CS98}.  The
combination AV18/Urbana is quite successful in describing the spectrum of nuclei up to
$A=8$~\cite{Pudliner:1997ck,Wiringa:2000gb}.  But, it does fail to predict the correct
isospin dependence of binding in these systems, and also underpredicts the spin-orbit
splitting of, e.g., the 3/2$^-$ and 1/2$^-$ resonances in the A=5 system.  Consequently,
the Urbana 3NF has been updated to produce a set of ``Illinois" potentials, which include
(phenomenologically, at least) the effect of ``pion ring" diagrams, and have 2--3
parameters that are tuned to reproduce levels in the spectra of nuclei up to $A \leq
8$~\cite{Pi01}. These potentials, when acting in concert with the AV18 NN force, do a good
job of describing spectra in systems with A=9 and 10~\cite{Pieper:2002ne}.

However,  it is not obvious that the Urbana and Illinois potentials are grounded in QCD.
Some of the structures are derived from diagrams involving pion exchange, but the
coefficient functions in front of those structures are, in some cases, chosen for ease of
numerical implementation, and given strengths which are adjusted to reproduce data. Closer
connection to the chiral symmetry of QCD was sought in, e.g. the Tucson-Melbourne 3N
potential, which considered the role of the $\rho$ meson, as well as the constraints of
chiral symmetry on the $\pi$N amplitude which appears in the two-pion-exchange
3NF~\cite{Co79,CP93}. The Brazilian 3NF also attempted to impose constraints from
chiral symmetry~\cite{RIF84}. 

The advent of chiral perturbation theory ($\chi$PT) as a tool for analyzing nuclear forces
resulted in the derivation of a 3NF which is in accord with the pattern of chiral-symmetry
breaking in QCD~\cite{vK94}. If the chiral expansion is applied directly to the 3N
potential---as was done in Ref.~\cite{vK94}---then three contributions occur at leading
order (LO). They are: a short-range, spin-isospin independent piece (as in the $V_{ijk}^R$
of Eq.~(\ref{eq:Urbana})); a piece associated with the short-range emission of a pion by
an NN pair with its subsequent absorption by the third nucleon; and a two-pion-exchange
3NF. The $\pi$N amplitude that appears in the two-pion-exchange piece of the chiral 3NF
involves LECs from ${\cal L}_{\pi N}^{(2)}$: $c_1$, $c_3$, and $c_4$. The LECs
$c_3$ and $c_4$ encode p-wave $\pi$N scattering, so $\chi$PT has the Fujita-Miyazawa force
as one of the dominant pieces of its 3NF. (Indeed, if a variant of $\chi$PT with an
explicit Delta degree of freedom is employed then the Fujita-Miyazawa 3NF occurs one order
earlier than the other pieces of the chiral 3NF~\cite{vK94,EKM08}.) The leading $\chi$PT
3NF has been used to investigate scattering in the 3N system~\cite{Ep02}, and nuclear
spectra in {\it ab initio} calculations up to A=13~\cite{Na07,Ma12}. And, as mentioned
above, it has, under certain approximations to the many-body physics, been shown to
improve descriptions of the binding of neutron-rich
nuclei~\cite{Otsuka:2009cs,Holt:2010yb}.  It has also been applied to obtain an equation
of state for neutron-rich matter~\cite{HS10}. 

In spite of these successes, puzzling discrepancies between theory and data persist. One
example is the analyzing power $A_y$ in neutron-deuteron scattering at low energies, with
a similar issue also occurring for neutron-${}^3$He scattering (see, e.g.
Ref.~\cite{Clegg:2009zz}). No modification of NN potentials which is consistent with the
NN data and the dominance of one-pion exchange at long range seems able to explain this
discrepancy, leaving the ``$A_y$ puzzle" firmly in the realm of 3NFs to resolve.  However,
neither the model 3NFs on the market, nor the LO chiral 3NF described in the previous
paragraph, can do so. Of course, extending a $\chi$PT calculation of the 3NF to higher
orders in the chiral expansion might reveal the operator and mechanism (or mechanisms)
which solves this problem, and work along these lines is in
progress~\cite{Be08,Be11,Krebs:2013kha,Girlanda:2011fh}. But, as the chiral order increases, classifying the
possible 3NF operators becomes very involved. It would be interesting to have an
additional tool that could help sort out the most relevant operator structures.

The $1/N_c$ expansion of QCD can be used to provide this kind of
insight~\cite{'tHooft:1973jz,Witten:1979kh}.  This approach
to the non-perturbative regime of QCD has proven very useful in the study of 
baryons \cite{Jenkins:2009wv}, for reviews see \cite{Jenkins:1998wy,Manohar:1998xv}.  In
the context of nuclear forces the $1/N_c$ expansion was first used to study the central
part of the NN potential by Savage and Kaplan~\cite{KS96}, and then to analyze the
complete potential, classifying the
relative strengths of the central, spin-orbit and tensor forces, by Kaplan and Manohar~\cite{KM97}. These authors analyzed the NN potential for
momenta of order $N_c^0$, i.e. $p \sim \Lambda_{QCD}$, and found that it is an expansion
in $1/N_c^2$. Furthermore, the $1/N_c^2 \approx 1/10$ (in our world) hierarchy between the
different contributions to the NN potential  is roughly borne out in the
Nijm93~\cite{St94} NN potential. The arguments that lead to this conclusion will be
recapitulated in Sec.~\ref{sec-NN}.

In this work we extend that analysis to the three-nucleon system, classifying the possible
operator structures that can contribute to a general 3NF according to a counting in
$1/N_c$. We do this by computing the energy of the 3N system as $N_c \rightarrow
\infty$, starting with the Hartree expansion for the nuclear Hamiltonian in the
large-$N_c$ limit~\cite{Dashen:1994qi,KM97}:
\begin{eqnarray}
H = N_c \sum_{s, t, m} v_{stm} \left(\frac{S}{N_c}\right)^s  
\left(\frac{I}{N_c}\right)^t \left(\frac{G}{N_c}\right)^m \ , 
\label{eq:Hartree}
\end{eqnarray}
where we suppressed spin and isospin indices in the spin-flavor structures $O=\{S,I,G\}$
and vector indices in the coefficients $v$. These coefficients are, in fact, ${\cal O}(1)$
functions of the momenta.  The explicit factors of $1/N_c$ ensure that an $m$-body
interaction scales generically as $1/N_c^{m-1}$, as mandated by large-$N_c$ QCD
counting~\cite{Witten:1979kh}.  Spin, isospin and vector indices are contracted so that
$H$ is rotation and isospin invariant, as well as parity even and time-reversal even.  In a
quark-operator basis the spin-flavor structures are given by one-body operators
\begin{eqnarray} \label{qop}
S^i = q^\dagger  \frac{\sigma^i}{2} q , \quad   I^a = q^\dagger  \frac{\tau^a}{2} q , \quad  G^{ia} = q^\dagger  \frac{\sigma^i \tau^a}{4} q ,  
\end{eqnarray}
where $q^\dagger, q$ are creation and annihilation operators for the light quarks $u,d$
and $\sigma$, $\tau$ are the standard $SU(2)$ Pauli matrices acting on spin and isospin,
respectively. Taken together, the 15 operators in Eq.~(\ref{qop}) generate the $SU(4)$
algebra
\begin{eqnarray}
[S^i,S^j]=i\epsilon^{ijk} S^k \ ,  && \ [S^i,G^{ja}]=i\epsilon^{ijk} G^{ka}, \nonumber \\
{[}I^a,I^b]=i\epsilon^{abc} I^c \ , && \ {[}I^a,G^{ib}]=i\epsilon^{abc} G^{ic},  \nonumber \\
{[}S^i,I^a]=0 \ , && {[}G^{ia},G^{ib}]=\frac{i}{4} \delta^{ij} \epsilon^{abc} I^c + \frac{i}{4} \delta^{ab} \epsilon^{ijk} S^k.
\label{eq:algebra}
\end{eqnarray}
Since we are interested in taking matrix elements between nucleon states we will indicate
with $O_{\alpha}$ that the operator $O$ acts on nucleon $\alpha=1,2,3$, so that $S,I,G$
in Eq.~(\ref{eq:Hartree}) can be any of $S_{\alpha},I_{\alpha},G_{\alpha}$.  But products of operators acting on the same nucleon in
Eq.~(\ref{eq:Hartree}) must be reduced to a single operator.  As is explained in Secs.~\ref{sec-NN} and
~\ref{sec-redrules}, this is achieved using the relations and reduction rules for the
powers of the basic operators ${S,I,G}$ that act on the same nucleon, which are discussed
in Ref.~\cite{Dashen:1994qi}.  The contributions to the 3NF that result after such
reduction can be straightforwardly estimated, since matrix elements of $S$ and $I$ between
nucleon states are ${\cal O}(1)$, which is in contrast to matrix elements of $G$, which
are ${\cal O}(N_c)$.  The leading force will thus be constructed out of $G$'s and unit
operators, acting on the different nucleons. In fact, the algebra Eq.~(\ref{eq:algebra})
was derived in the one-nucleon sector for external nucleon momenta of order $N_c^0$, and so this conclusion holds in
that kinematic regime (a similar remark applies to the NN potential derived in
Ref.~\cite{KM97}). If results for lower momenta are desired then the counting of operators
obtained here can be modified accordingly. We present the analysis of leading and
sub-leading 3NFs in the $1/N_c$ expansion in  Sec.~\ref{sec-3N}, and summarize our
conclusions in Sec.~\ref{sec-conclusion}. 

In the large-$N_c$ limit the mass of the nucleon tends to infinity. This provides both a
problem and an opportunity for computation of the nuclear potential. The opportunity
arises because, in this limit, the nuclear potential can be computed as the static 
energy of the system in a fixed configuration in co-ordinate space (for analogous studies
of heavy-quark systems on the lattice see Ref.~\cite{NPLQCD}). This implies that the 3N
potential (modulo issues of exchange diagrams, see below) obtained from our argument is
local, being, e.g. a function of the Jacobi co-ordinates $r_{12}$ and $r_3$
(velocity-dependent forces arise at sub-leading orders in $1/N_c$, and lead to
non-localities). The problem exists because the only measurable quantity in this
infinitely-massive-nucleon limit is the total potential energy, and the large-$N_c$
analysis gives no information on the dependence of the force on $r_{12}$ and $r_3$---at
least none beyond the statement that the function encoding that dependence has a size
given by $N_c$ counting. Thus, since we only ``measure" the total potential energy, and we
cannot tell which pieces depend only on, say, $r_{12}$, we can make no {\it a priori}
distinction between contributions to that energy from NN interactions, and contributions
from 3NFs. The best we can do is to identify operator structures which occur in the 3N
energy, and do not arise within the large-$N_c$ analysis of the NN potential of
Ref.~\cite{KM97}. 

One might be concerned that a 3NF derived from large-$N_c$ cannot be in accord with the
meson-exchange picture used successfully for many years to derive NN and 3N forces.  In
Refs.~\cite{Ba02,BC02} Banerjee {\it et al.} and Belitsky and Cohen explored the
relationship between this picture of the nuclear force and the large-$N_c$ analysis of
Ref.~\cite{KM97}.  Initially it appeared that multi-meson-exchange graphs led to
violations of the large-$N_c$ scaling of the NN potential: in particular to pieces of the
NN potential that scaled with powers of $N_c$ larger than one. However, Ref.~\cite{Co03}
later explained this apparent discrepancy between the meson-exchange and large-$N_c$
pictures by pointing out that the potentials analyzed in Refs.~\cite{Ba02,BC02} were
energy dependent, whereas almost all NN interactions used for phenomenological purposes
are energy independent. Ref.~\cite{Co03} concluded that an energy-independent NN potential
could have $N_c$ scaling consistent with that derived in Ref.~\cite{KM97}, and so
large-$N_c$ analysis is not inconsistent with a meson-exchange picture of nuclear forces
for the NN case. An important point for a successful matching calculations is that the
Hartree Hamiltonian Eq.~(\ref{eq:Hartree}) and the $SU(4)$ algebra Eq.~(\ref{eq:algebra})
implicitly assume the presence of the $\Delta$ resonance with $S=I=3/2$. In our discussion
of the NN and NNN potentials, when taking matrix elements, we project $H$ to the nucleons-only
piece of the Hilbert space.  We have not performed a matching calculation to check the consistency with the
meson-exchange picture for the 3N potential, but it would
be a worthy subject for future study.

One might also wonder whether double counting will result if the 3N potential obtained
from the large-$N_c$ analysis is used in a multi-nucleon Schr\"odinger equation. To address
this issue we note that another assumption made in the derivation of the algebra
Eq.~(\ref{eq:algebra}) was that meson energies are of order $\Lambda_{QCD}$. This implies that
the energy of the intermediate nucleon state in the 3NF (see, e.g., Fig.~\ref{fig:Vdir})
must be order $\Lambda_{QCD}$ if an analysis based on this algebra is to prevail. Having
states of this energy included in the computation of the nuclear potential is consistent
with the insertion of the resulting nuclear force in the 3N Schr\"odinger equation (or,
equivalently, a Faddeev equation) provided a momentum cutoff is employed there. If that
momentum cutoff is above $\Lambda_{QCD}$, but below $\sqrt{N_c} \Lambda_{QCD}$, the
intermediate nucleonic states with energies of order $\Lambda_{QCD}$ (i.e. momenta $\sim
\sqrt{M \Lambda_{QCD}}$) will not be accounted for by the iteration of the potential via
the Schr\"odinger/Faddeev equation, and so should be included in the potential. The NN and
3N interactions derived here, and in Ref.~\cite{KM97}, can thus be inserted into the
quantum-mechanical equation and used to compute the wave function of nuclear systems.
 
With the conceptual underpinning of a 3NF in large-$N_c$ QCD defined, and the
circumstances under which it should be used in a Schr\"odinger equation for a
multi-nucleon system clarified, we now turn back to the NN system, in order to explain how
the corresponding analysis works in that, simpler, case.
 
\section{The NN potential in the $1/N_c$ expansion: review}

\label{sec-NN}

Here we review the $1/N_c$ analysis of Kaplan and Manohar~\cite{KM97} for the two-nucleon
potential, setting up the notation that we will use later in Sec.~\ref{sec-3N} to analyze
the three-nucleon force. In Ref.~\cite{KM97} the large-$N_c$ expansion was used to analyze
the object: 
\begin{equation}
U_{{\rm NN}}^A=(1-P_{12}) U \ ,
\end{equation} 
where $U$ is the sum of all direct diagrams, and $P_{ij}$  is the permutation operator
that switches {\it all} quantum numbers of particles $i$ and $j$.  In nuclear physics
computations it is the operator $U$ which is inserted into the Schr\"odinger equation. The
correct anti-symmetry properties of the nuclear state are then imposed by computing matrix
elements only in partial waves which are allowed by the Fermi-Dirac statistics of the
nucleons. 

In order to discuss the momentum dependence of the potential we first define initial and
final relative momenta:
\begin{equation}
{\bf p}={\bf p}_1 - {\bf p}_2 \ , \quad  {\bf p}'={\bf p}'_1 - {\bf p}'_2 \ , 
\end{equation}
where ${\bf p}_i ({\bf p}'_i)$ is the initial (final) momentum of the $i$-th nucleon.  To
simplify later analysis we also define time-reversal-odd (T-odd) and time-reversal-even
(T-even) combinations of these:
\begin{equation}
{\bf p}_{\pm}={\bf p}' \pm {\bf p} \ .
\label{ppm}
\end{equation}
Notice that ${\bf p}_{+}$ is T-odd and ${\bf p}_{-}$ is T-even, as initial and final
states are also exchanged under time reversal.  Both combinations, being vectors,  are odd
under parity.  In $U$ only ${\bf p}_-$ enters at leading order in $N_c$  since the
potential is local at this order in the $1/N_c$ expansion.  Powers of ${\bf p}_+$ indicate
the presence of non-locality. In a meson-exchange picture they arise due to the occurrence
of relativistic corrections suppressed by $1/M_N$. Thus, each appearance of a power of
${\bf p}_+$ costs a power of $1/N_c$.  Finally, energy conservation and the constraint
that the external NN states in a diagram be on-shell results in
\begin{equation}
{\bf p}_+ \cdot {\bf p}_-=0 \ ,
\label{eq:onshellconstraintNN}
\end{equation}
which allows to eliminate this momentum structure.  In Ref.~\cite{KM97} the potential $U$
was written as a sum of products of one-body operators, including the explicit factors of
$1/N_c$ as shown in the Hartree Hamiltonian, Eq.~(\ref{eq:Hartree}). Isospin invariance of
the interaction requires that all isospin indices are contracted.

In general, operators acting on the same nucleon with spatial or isospin indices
contracted can be simplified.  For instance, $G^{ia}G^{ia}$ can be reduced to the unit
operator and a subleading contribution using
\begin{equation}
G^{ia}G^{ia} = \frac{3}{16} N_c (N_c+4) \mathds{1} - \frac{1}{4} I^a I^a - \frac{1}{4} S^i
S^i \ , 
\label{eq:Casimir}
\end{equation}
which is obtained from the quadratic $SU(4)$ Casimir evaluated on the completely symmetric
representation $S_{N_c}$. 
If the spatial indices are not contracted we have the more general identity
\begin{equation}
G^{ia} G^{ja} = \frac{1}{16} N_c (N_c+4) \delta^{ij} \mathds{1} - \frac{1}{4} \delta^{ij} S^2 +
\frac{1}{4} S^i S^j + \frac{i}{4} \epsilon^{ijk} S^k.
\label{eq:GGred}
\end{equation}
The complete set of operator reduction rules can be found in \cite{Dashen:1994qi}.

It is thus sufficient to consider structures where the contracted indices are carried by
operators acting on different nucleons. For instance, the leading order of the angular
momentum zero ($L=0$) component of the potential is obtained from
\begin{eqnarray}
U_{L=0}^{N_c} &\subset&  
N_c \sum_{{n=0}}^{N_c} u_n({\bf p}_-^2) (N_c^{-2} G_1^{ia}  G_2^{ia})^{n} \ , 
\label{eq:NNLOGs}
\end{eqnarray}
where $u_n({\bf p}_-^2)$ are arbitrary scalar functions of ${\bf p}_-^2$ that scale like
${\cal O}(N_c^0)$. This yields two strings of $G$'s, one on each of the two nucleons, with no contracted 
indices
amongst the $G$'s which act on an individual nucleon. Each such string of $G$'s can, 
nevertheless, be reduced,  because the matrix element of a general $m$-quark operator between single-baryon states scales as
\cite{KS96,KM97,Dashen:1994qi}
\begin{eqnarray}
\langle {\rm B}_1 |N_c^{-m}  O_m  | {\rm B}_1 \rangle = \frac{1}{N_c^{|I-S|}} \ .
\label{Oscaling}
\end{eqnarray}
Therefore the dominant parts in the operator resulting from each string of $G$'s have $I=S$. If, in addition, we restrict ourselves
to the case that the baryon is a nucleon only $(I,S)=(0,0), (1,1)$
contribute.  But those $I=S=0$ and $I=S=1$ operators can, via the Wigner-Eckart theorem, be replaced by the ${\cal O}(N_c^0)$ one-body operators $\mathds{1}$
and $N_c^{-1} G^{ia}$, up to a proportionality constant that ultimately gets absorbed in the undetermined functions of momenta that
appear in the large-$N_c$ NN potential. Thus, on a single-nucleon state, each string of $G$'s with uncontracted indices yields a matrix element that can be written:
\begin{eqnarray}
<{\rm N} | \underbrace{G G  ... G } _r|{\rm N} >  = N_c^r < \mathds{1} > + N_c^{r-1} <G> + {\cal O}
(N_c^{r-2}),
\label{eq:manyG}
\end{eqnarray}
where the  spatial and isospin indices on the RHS of Eq,~(\ref{eq:manyG}) are carried by Kronecker
$\delta$'s and the completely antisymmetric tensor $\epsilon$. For an example see the
Appendix, in particular Eq.~(\ref{twoGred}).

Eqs.~(\ref{eq:manyG}) and (\ref{eq:GGred}) show that it is enough to consider the one-quark operators
$\mathds{1}$ and $N_c^{-1} G_{ia}$ acting within each nucleon to construct the leading-order
spin-flavor structures. With this simple rule one obtains correctly
the explicit $1/N_c$ suppression factors contained in the Hartree expression,
Eq.~(\ref{eq:Hartree}), for the NN interaction.

The leading-order spin-flavor structures are thus
$\mathds{1}_1 \mathds{1}_2$ and $G^{ia}_1  G^{ja}_2 $.  The next step is to  project
out the different spin components of the leading-order $G_1 G_2$ tensor, namely
\begin{equation} \label{LOstrucs}
\quad G_1^{ia} G_2^{ia}, \quad \epsilon^{ijk} G_1^{ja} G_2^{ka}, \quad \left[G_1^{ia} G_2^{ja}\right]_2 \ ,
\end{equation}
where the first two correspond to $S=0$ and $1$ respectively, and 
\begin{equation}
\left[G_1^{ia} G_2^{ja}\right]_2 \equiv  G_1^{ia}
G_2^{ja} + G_1^{ja} G_2^{ia} - \frac23 \delta^{ij} G^{ka}_1 G^{ka}_2
\label{eq:spintensor}
\end{equation}
is the $S=2$
component.   The final step is the reduction of the operator $G$ to $\sigma^i \tau^a$ when restricted
to the nucleon subspace. Table~\ref{table-NNops} shows a complete set of independent spin-flavor
structures in the NN subspace, together with their $1/N_c$ scalings, spin content and time-reversal properties.

Each of these spin-flavor structures must then be combined with 
tensors formed out of the momenta ${\bf p}_-, {\bf p}_+$
to form a T-even, P-even, rotationally invariant operator. 
In particular, the $S=2$ structure (\ref{eq:spintensor}) 
must be contracted with a spatial tensor of rank two. Since at LO we have a local NN potential 
the only possible LO tensor is ${\bf p}_-^i {\bf p}_-^j$. Meanwhile, the second
($S=1$) spin-flavor structure must be contracted with a three-vector. Parity
invariance suggests ${\bf p}_- \times {\bf p}_+$ is the only possible candidate. However,
${\bf p}_- \times {\bf p}_+$ is odd under time reversal. And the constraint
(\ref{eq:onshellconstraintNN}) means we cannot multiply by powers of the T-odd
rotational scalar ${\bf p}_+ \cdot {\bf p}_-$---at least not on-shell. Thus our $S=1$ spin-flavor
structure cannot be multiplied by any combination of three-vectors that
results in an overall P-even, T-even object. The operator $\epsilon^{ijk} G_1^{ja}
G_2^{ka}$ will therefore not appear in the parity-conserving, time-reversal-non-violating
NN force~\cite{Wolfenstein}.  Finally, the first structure in Eq.~(\ref{LOstrucs}) and the unit operator
are the two leading-order $S=0$, $L=0$ operators.

\begin{table}
\begin{tabular}{cc|cc|cc}
\hline \hline
 $O$ & \ \  Order \ \  & $O_{\tau \tau}$  & \ \ Order \ \   & \ $S$ \ &  T \\
\hline
$ \mathds{1} $                           & 1         & $\tau_1 \cdot \tau_2$                                & $1/N_c^2$ &  0  &     +   \\ 
$\sigma_1 \cdot \sigma_2$       & $1/N_c^2$ & $\sigma_1 \cdot \sigma_2 \ \tau_1 \cdot \tau_2$      &       1   &  0  &     +   \\ 
$\sigma^i_1$                    & $1/N_c$   & $\sigma^i_1 \ \tau_1 \cdot \tau_2$                   & $1/N_c$   &  1  & \bf{--}  \\ 
$\sigma^i_2$                    & $1/N_c$   & $\sigma^i_2 \ \tau_1 \cdot \tau_2$                   & $1/N_c$   &  1  & \bf{--}  \\ 
$(\sigma_1 \times \sigma_2)^k $ & $1/N_c^2$ & ($\sigma_1 \times \sigma_2)^k \ \tau_1 \cdot \tau_2$ & $1$       &  1  &  + \\ 
$[\sigma^i_1 \sigma^j_2]_{_2} $ & $1/N_c^2$ & $[\sigma^i_1 \sigma^j_2]_{_2} \ \tau_1 \cdot \tau_2$ & $1$       &  2  &  + \\ 
\hline
\hline
\end{tabular}
\caption{Spin-flavor structures for the two-nucleon potential. The $(\sigma_1 \times \sigma_2)$ structure arises in the large-$N_c$
analysis, but its appearance in $U$ is
 precluded by permutation
symmetry.}
\label{table-NNops}
\end{table}

The rotational scalars formed in this way may always be multiplied by
an arbitrary scalar function of ${\bf p}_-^2$. Therefore, to leading order
\begin{equation} \label{eq:orderNc}
U^{N_c} = N_c \left( \ U^1_{S}({\bf p}_-^2) \mathds{1} + U^2_{S}({\bf p}_-^2) \ \sigma_1 \cdot \sigma_2 \ \tau_1 \cdot \tau_2 
+ U^1_{D}({\bf p}_-^2) \ [{\bf p}_- {\bf p}_-]_{_2} \cdot [\sigma_1 \sigma_2]_{_2} \ \tau_1 \cdot \tau_2 \ 
\right) \ ,
\end{equation}
with 
\begin{equation}
[A_i B_j ]_{_2} \equiv A_i B_j + A_j B_i - \frac23 \delta_{ij} A\cdot B
\label{eq:rank2tensor}
\end{equation}
 the $L=2$ component of 
the tensor $A_i B_j$ constructed out of two vector quantities, and $U_S^{1,2}({\bf p}_-^2)$, $U_D^1({\bf p}_-^2)$ arbitrary ${\cal O}(1)$ scalar functions of ${\bf p}_-^2$.
As discussed above, there are no $S=1$ terms at leading order. 

Sub-leading corrections are associated with
$1/N_c$-suppressed operators.  Such suppression may occur for two reasons. Firstly, NN
operators involving $S$ and $I$, instead of $G$, will be reduced by factors of $1/N_c$,
because of the $N_c$-scaling of the nucleonic matrix elements of these operators. The
second source of $1/N_c$ suppression is the appearance in expressions of the momentum
${\bf p}_+$.  Time reversal and parity conservation conspire so that the expansion is in
$1/N_c^2$.

With these two results regarding $1/N_c$ suppression in hand, Kaplan and Manohar concluded
that the following operators give contributions to the NN potential of ${\cal
O}(1/N_c)$ (see also Table~\ref{table-NNops}):
\begin{eqnarray}
U^{1/N_c} &=& \delta^{(2)} U^{N_c} + N_c^{-1} \left( \ U^3_{S} \ {\bf p}_+^2 \ \mathds{1} + 
\ U^4_{S} \ \sigma_1 \cdot \sigma_2 
+ U^5_{S} \ \tau_1 \cdot \tau_2 
+ U^6_{S} \ {\bf p}_+^2 \ \sigma_1 \cdot \sigma_2 \ \tau_1 \cdot \tau_2 
\right.  \nonumber \\
&& 
\qquad \qquad \qquad \quad + \ U^1_{P} \ ({\bf p}_+ \times {\bf p}_-) \cdot ({\bf \sigma}_1 + {\bf \sigma}_2) 
 +  U^2_{P} \ ({\bf p}_+ \times {\bf p}_-) \cdot ({\bf \sigma}_1 + {\bf \sigma}_2) \tau_1 \cdot \tau_2 \nonumber \\
&& \left.
\qquad \qquad \qquad \quad + \ U^2_{D} \ [{\bf p}_- {\bf p}_-]_{_2} \cdot [\sigma_1\sigma_2]_{_2} \ 
+ U^3_{D} \ [{\bf p}_+ {\bf p}_+]_{_2} \cdot [\sigma_1 \sigma_2]_{_2} \ \tau_1 \cdot \tau_2 \ 
\right) \ .
\label{eq:order1overNc}
\end{eqnarray}
At this order the leading-order operators appear again, as they can also
be obtained by replacing one $G^{ia}/N_c$ by $ (S^i I^a)/N_c^2$ in the Hartree
Hamiltonian. We denoted this contribution by $\delta^{(2)} U^{N_c} $ in the expression above.
The spin-flavor structures that appear here (and in Table~\ref{table-NNops}) and momentum tensors with up to four momenta can also be read off from the
results for the 3NF that will be presented later in Sec.~\ref{sec-3N},
by eliminating the third nucleon and only keeping momentum structures that depend on ${\bf p}_\pm$.
Here we only show the potential up to quadratic structures in momenta (modulo arbitrary functions 
of ${\bf p}_-$).

Comparing Table~\ref{table-NNops} with Eq.~(\ref{eq:orderNc}) and Eq.~(\ref{eq:order1overNc}) one can 
see that the spin-flavor structures proportional to  $\sigma_1
\times \sigma_2$ are missing because, as discussed for the LO case, they need to be multiplied by a T-even, P-even, $L=1$ 
momentum structure, which cannot be constructed in the NN case.

However, there is an additional constraint from permutation symmetry \cite{Wolfenstein}.
For example, $\sigma_1, \sigma_2$ only appear in the  $\sigma_1 +  \sigma_2$ combination.
The  $\sigma_1 -  \sigma_2$ combination is excluded by permutation symmetry, as it is
T-odd and parity-even  and needs to be contracted with a vector built from  $\mathbf{p}_+,
\mathbf{p}_-$, where  $\mathbf{p}_\pm$ are both odd under exchange of the nucleons $1,2$.
For instance, if we would start from the general structure 
\begin{eqnarray}
U(\mathbf{p}_-^2) (\mathbf{p}_+ \times \mathbf{p}_-) \sigma_{1} + U'(\mathbf{p}_-^2) (\mathbf{p}_+ \times \mathbf{p}_-) \sigma_{2} \ 
\end{eqnarray}
permutation symmetry imposes $U=U'=U_P^1$ so that only the symmetric spin-flavor structure
$\sigma_1+\sigma_2$ appears in Eq.~(\ref{eq:order1overNc}). The $\sigma_1 \times \sigma_2$ structure can
also be eliminated by permutation symmetry.

In summary, to leading order (${\cal O}(N_c)$) there are two structures with $L=0$ and one
with $L=2$. To subleading order (${\cal O}(1/N_c)$) and up to two momenta,  there are four
structures with $L=0$, two with $L=1$ and two with $L=2$.

This translates into definite scaling predictions for the different parts of the NN
potential, which in the usual form is given by
\begin{eqnarray} \label{VNNpot}
V_{NN}&=&V^0_C+V^0_{SS} \ S_1 \cdot S_2 + V^0_{LS} \ L\cdot S + V^0_T \ S_{12} + V^0_Q \
Q_{12} \nonumber \\
&& + (V^1_C+V^0_{SS} \ S_1 \cdot S_2 + V^1_{LS} \ L\cdot S + V^1_T \ S_{12} + V^1_Q \
Q_{12}) \ \tau_1 \cdot \tau_2 \ . 
\end{eqnarray}
Here $L$ is the angular momentum operator, which is T-odd and P-even and in our notation 
is replaced via the Wigner-Eckart theorem by the ${\bf p}_+ \times {\bf p}_-$ structure.
The quadratic spin-orbit interaction $Q_{12}$  involves four momenta in our 
notation and we did not include it in Eq.~(\ref{eq:order1overNc}).

A comparison with ``experiment" can be achieved by comparing with a successful
phenomenological potential. This has been done in Ref.~\cite{KM97} using the Nijmegen
potential~\cite{St94}. The $1/N_c$ scaling of the different structures in
Eq.~(\ref{VNNpot}) translates into a hierarchy for the functions used to parametrize the  Nijmegen
potential, which is well satisfied by their numerical values, as discussed in detail in 
Ref.~\cite{KM97}.
 
Although in the two-nucleon case the operator structure of the interaction is simple
enough to be obtained by explicit construction, as sketched above, at this point it is
useful to discuss a more systematic way of counting the number of spin-flavor structures
that can contribute, something that will prove very useful in the more involved
three-nucleon case. The systematic classification can be done as follows.
 
The number of independent spin-flavor structures $O_{IS}$ of isospin $I$ and spin $S$ that
can contribute to the matrix element $\langle {\rm NN} | O_{IS} | {\rm NN}   \rangle $ can
be obtained by considering the decomposition of $\overline{\mathbf{R}} \otimes
\mathbf{R'}$, with $\mathbf{R},\mathbf{R'}$ the irreducible representations of
spin-flavor for the two nucleons, so that the matrix element is a scalar. To obtain the
possible irreps $\mathbf{R}$ we decompose the tensor product of two-nucleon states, each
nucleon transforming as the fundamental representation of $SU(4)$   
\begin{eqnarray}
\stackrel{\mathbf{4}}{\raisebox{-0.1cm}{\yng(1)}} &\equiv& \{ p \uparrow, p \downarrow,  n \uparrow, n \downarrow \} \ .
\end{eqnarray}
The two-nucleon states are obtained as the decomposition of the tensor product
$\mathbf{4} \otimes \mathbf{4} = \mathbf{6} \oplus  \mathbf{10}$. In terms of Young tableaux  
\begin{eqnarray}
\stackrel{\mathbf{4}}{\raisebox{-0.1cm}{\yng(1)}} \otimes \stackrel{\mathbf{4}}{\raisebox{-0.1cm}{\yng(1)}}
&=& 
\stackrel{\mathbf{6}}{\raisebox{-0.5cm}{\yng(1,1)}} \oplus 
\stackrel{\mathbf{10}}{\raisebox{-0.1cm}{\yng(2)}} \ . 
\end{eqnarray}
As states and operators are labeled by their  isospin and spin transformation properties,
we decompose  $SU(4)$ irreps in $SU(2)_I \times SU(2)_S \subset SU(4)$, labeled by $(2 I +
1, 2 S + 1)$. The result is (only $SU(4)$ irreps are in boldface):
\begin{eqnarray}
\stackrel{\mathbf{10}}{\raisebox{-0.1cm}{\yng(2)}} &=& 
\left(\stackrel{3}{ \raisebox{-0.1cm}{\yng(2)}} \ , \ \stackrel{3}{\raisebox{-0.1cm}{\yng(2)}} \right)  \oplus  
\left( \stackrel{1}{ \raisebox{-0.5cm}{\yng(1,1)}} \ , \
\stackrel{1}{\raisebox{-0.5cm}{\yng(1,1)}} \right) \ , \\
\stackrel{\mathbf{6}}{\raisebox{-0.5cm}{\yng(1,1)}} &=& 
\left(\stackrel{3}{ \raisebox{-0.1cm}{\yng(2)}} \ , \ \stackrel{1}{\raisebox{-0.5cm}{\yng(1,1)}} \right)  \oplus  
\left(\stackrel{1}{ \raisebox{-0.5cm}{\yng(1,1)}} \ , \
\stackrel{3}{\raisebox{-0.1cm}{\yng(2)}} \right)  \ . 
\end{eqnarray}
With this result in hand we can determine the number, and type, of spin-flavor structures that
occur in $O_{IS}$. 
We consider the decomposition of $\overline{\mathbf{R}} \otimes \mathbf{R'}$, with
$\mathbf{R},\mathbf{R'}=\mathbf{6},\mathbf{10}$ into irreps of $SU(2)_I \times SU(2)_S$.
We are interested in the pieces of the direct product that yield $I=0$ operators, which are:
\begin{eqnarray}
\sum_\mathbf{R,R'} \overline{\mathbf{R}} \otimes \mathbf{R'} &\supset& 4 (0,0) \oplus 6 (0,3) \oplus 2 (0,5) + \dots,
\end{eqnarray}
i.e. the direct product contains four independent isoscalar structures of $S=0$, six of $S=1$ and
two  of $S=2$.  Their explicit forms are $\mathds{1}, \sigma_1 \cdot
\sigma_2$, $\sigma_1, \sigma_2, \sigma_1 \times \sigma_2$ and $[\sigma_1 \sigma_2]_{_2}$,
each of which can be multiplied by any of the two isospin invariants
$\mathds{1},\tau_1 \cdot \tau_2$. The resulting spin-flavor structures are collected in Table~\ref{table-NNops}.

This finishes the review of the NN case. We proceed now to the construction of the 3N
potential.

\section{The 3N potential in the $1/N_c$ expansion}
\label{sec-3N}

In this Section we will extend the analysis that we reviewed for the NN potential to the
case of the 3N potential. The sum of all  3N$\rightarrow$3N diagrams can be written in operator form as:
\begin{equation}
V_{\rm 3N}^A=(1 + P_{12} P_{23} + P_{13} P_{23})(1-P_{23}) V \ .
\label{eq:DplusE}
\end{equation}
The terms in parentheses in Eq.~(\ref{eq:DplusE}) thus generate the exchange diagrams
necessitated by the identicality of the nucleons from the operator $V$, which itself is
the sum of all direct diagrams (see, e.g., Fig.~\ref{fig:Vdir}), and is the object that enters the
Schr\"odinger equation in nuclear-physics computations. We will classify the
structures that contribute to $V$, and derive their scaling behaviour with
$N_c$.

We do this by first discussing the momenta involved, and the possible momentum
structures obtainable therefrom. We then derive the LO spin-flavor structures, 
and count all possible spin-flavor structures. We finish with the explicit construction of the
operators, including the spatial part.

\subsection{Momenta and momentum structures}

Throughout, we work in the 3N center-of-mass frame, where:
\begin{equation}
{\bf p}_1 + {\bf p}_2 + {\bf p}_3={\bf p}_1' + {\bf p}_2' + {\bf p}_3'=0 \ . 
\end{equation}
Any graph can then be expressed as a function of the Jacobi momenta ${\bf p}$ and ${\bf q}$ 
\begin{equation}
{\bf p}={\bf p}_1 - {\bf p}_2 \ , \qquad  {\bf q}={\bf p}_3 - \left({\bf p}_1 + {\bf
p}_2\right)/2 \ ,
\end{equation}
so that the 3NF can be written as a function of
four three-momenta ${\bf p}$, ${\bf p}'$, ${\bf q}$, and ${\bf q}'$. 

\begin{figure}[t] 
   \centering
$
\begin{array}{c}
 \includegraphics*[width=10cm]{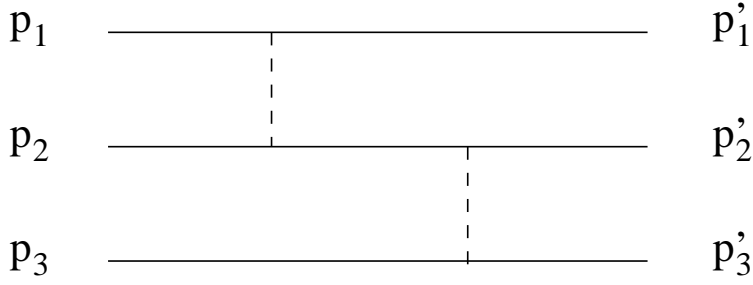}
\end{array}
$
\caption{Tree level two meson exchange contribution to the 3NF. }
   \label{fig:Vdir}
\end{figure}

Conservation of energy yields the constraint:
\begin{equation}
{\bf p}^2 + \frac43 {\bf q}^2={\bf p}^{\prime \, 2} + \frac{4}{3} {\bf q}^{\prime \, 2}.
\label{eq:onshell}
\end{equation}
In terms of the momenta with well-defined properties under time reversal, ${\bf p}_{\pm}$
as in Eq.~(\ref{ppm}), and the analogous ${\bf q}_{\pm}={\bf q}' \pm {\bf q}$, the
constraint (\ref{eq:onshell}) becomes
\begin{equation}
{\bf p}_+ \cdot {\bf p}_-=-\frac{4}{3} {\bf q}_+ \cdot {\bf q}_- \ ,
\end{equation}
which will allow us to eliminate ${\bf q}_+ \cdot {\bf q}_-$ in favor of ${\bf p}_+ \cdot
{\bf p}_-$.  Analysis of the contributions to Fig.~\ref{fig:Vdir} shows that the presence
of $ {\bf p}_+ ,  {\bf q}_+$ comes from relativistic corrections that introduce powers of
$1/M_N$, so that each power of either $ {\bf p}_+$  or  $ {\bf q}_+$ is associated with a
supression factor of $1/N_c$. The LO momentum structures are ${\cal O}(N_c^0)$ and depend
only on  $ {\bf p}_-$  and  $ {\bf q}_-$. They correspond to local potentials. 

In fact, since all spin-flavor structures are built out of P-even objects, parity invariance of $V$ 
requires that the momentum structures appearing---both at leading and sub-leading orders in $1/N_c$---
must contain an even number of momenta. 
In Table~\ref{table-momenta} we show the TP properties, $L$ content and order in $1/N_c$ of
the 3N-system momentum tensors which contain up to two momenta. Time-reversal-odd momentum structures only appear at subleading orders, 
as they must include at least either a ${\bf
p}_+$ or a ${\bf q}_+$.

\begin{table}
\begin{tabular}{cccccc}
\hline \hline  
\ \ & T & P & $\Pi^{\rm{TP}}$ & $L$ & Order\\
\hline 
&+ & \bf{--} & ${\bf p}_-$ & 1 & 1 \\
&+ & \bf{--} & ${\bf q}_-$ & 1 & 1 \\
& \bf{--} & \bf{--} & ${\bf p}_+$ & 1 & $1/N_c$ \\
&\bf{--} & \bf{--} & ${\bf q}_+$ & 1 & $1/N_c$ \\
& \bf{--} & + & ${\bf p}_{+} {\bf p}_{-}$  & \ \ 0,1,2 \ \ & $1/N_c$\\
& \bf{--} & + & ${\bf q}_{+} {\bf q}_{-}$ & 0,1,2 & $1/N_c$\\
& \bf{--} & + & ${\bf p}_{\pm} {\bf q}_{\mp}$ & 0,1,2 &$1/N_c$, $1/N_c$ \\
& + & + & ${\bf p}_{\pm} {\bf q}_{\pm}$ & 0,1,2 &$1/N_c^2$, $1$ \\
& + & + & ${\bf p}_{\pm} {\bf p}_{\pm}$  & 0,2  &$1/N_c^2$, $1$ \\
& + & + & $ \ \ \ {\bf q}_{\pm} {\bf q}_{\pm} \ \ \ $  & 0,2  & \ \ \ \ $1/N_c^2$, $1$ \ \ \ \ \\
\hline \hline 
\end{tabular}
\caption{Tensors $\Pi^{\rm{TP}}$ constructed from up to two 3N-system momenta, together
with their $\rm{T}$ and $\rm{P}$ properties and their angular momentum ($L$)
content.  Note that in the NN system none of the tensors involving ${\bf q}_{\pm}$
are present.  The $\pm$ signs in the subscripts are always to be read as
correlated, so that the last four entries in the Table each contain two
possible tensors. The last column shows the $1/N_c$ order at which the corresponding 
momentum structure appears. }
\label{table-momenta}
\end{table}

\subsection{Leading spin-flavor structures}
\label{sec-redrules} 

As in the NN case, leading-order spin-flavor structures are obtained from products of an arbitrary number of $G$'s, with 
their indices contracted in order to get isoscalar operators of spin $S=0,1,2,3$, which are the only quantum numbers 
relevant for isospin conserving interactions in the 3N subspace.   Any spatial indices associated with the spin tensor of rank-0,1,2,3 
are then
contracted with a momentum tensor of the same rank to form a singlet, so that the interaction
is invariant under rotations. 

For example, the leading order $S=0$ 3N structures are obtained from
\begin{eqnarray}
V_{L=0}^{N_c} &\subset&  
N_c  \sum_{{n_{12},n_{13},n_{23}} \atop n_{123}}  v_{{n_{12},n_{13}\dots}} (N_c^{-2}
G_1^{ia}  G_2^{ia})^{n_{12}} (N_c^{-2} G_1^{jb}  G_3^{jb})^{n_{13}}  \nonumber \\  
& & \qquad \qquad \qquad \qquad \quad  \times            (N_c^{-2} G_2^{kc}  G_3^{kc})^{n_{23}} 
 (N_c^{-3} \epsilon^{lmr} \epsilon^{def}  G_1^{ld} G_2^{me} G_3^{rf} )^{n_{123}} \nonumber \\
& & \qquad \qquad \qquad + \cdots 
\label{eq:LOGs}
\end{eqnarray}
where the dots stand for terms with more complex index contractions. A general structure
has the form $O_\alpha O_\beta O_\gamma$, with the greek index indicating the nucleon on
which a particular $O$ acts.  As in the NN case, products like $G_1^{ia} G_1^{ja}$ where there is at
least one index contracted between operators acting on the same nucleon are not included.
The structures shown in Eq.~(\ref{eq:LOGs}) still seem hard to reduce, but this
can be achieved after taking matrix elements in the NNN subspace using
Eq.~(\ref{Oscaling}) and Eq.~(\ref{eq:manyG}), as we did in the NN case.  The simple rule is
again that, at leading order, an arbitrary product of $G$'s can be reduced to a sum of $I=S$
operators, which for the N subspace reduce just to the unit operator and one $G$. So, as in the NN force, 
the LO structures are found by considering one-quark operators $\mathds{1}$ and $N_c^{-1} G^{ia}$ acting on
each nucleon. This gives the explicit $1/N_c$ suppression factors that come from the spin-flavor
part.  Then, within the N subspace we replace $G^{ia}$ by
$\sigma^i \tau^a$. Bearing in mind that spin and isospin indices should be contracted with $\delta_{ij},\delta_{ab}$ or
$\epsilon_{ijk},\epsilon_{abc}$ tensors one straightforwardly obtains the leading spin-flavor structures shown in
Table~\ref{table-leadingops}. 

The isospin structures are the unit operator, the three scalar
products $\tau_\alpha \cdot \tau_\beta$ and a new structure that was not present in the NN
case, the triple product $(\tau_\alpha \times \tau_\beta) \cdot \tau_\gamma$. It is
important to notice that the triple product of $\tau$ is time-reversal odd, as under time
reversal $(\tau^1, \tau^2, \tau^3) \rightarrow (\tau^1, -\tau^2, \tau^3)$. This is in contrast to 
$(\sigma^1, \sigma^2, \sigma^3) \rightarrow (-\sigma^1, -\sigma^2, -\sigma^3)$. The
different transformation properties of the spin and isospin operators under time reversal
just reflect the fact that under time reversal spins get flipped, while protons and
neutrons retain their identity and are not exchanged. 

\begin{table}
\begin{tabular}{lccccc}
\hline \hline
Spin content &  & \   & LO quark operator & $\sigma\tau$-projection & Multiplicity \\
\hline
$S=0$ &   &   & $\mathds{1}$ & $\mathds{1}$ & 1 \\ 
             &   &   & $ N_c^{-2} \ G_\alpha^{ia} G_\beta^{ia} $ 
                     & $ \sigma_\alpha \cdot \sigma_\beta \ \tau_\alpha \cdot \tau_\beta $ & 3 \\
             &   &   & $ N_c^{-3} \ \epsilon^{ijk}\epsilon^{abc} G^{ia}_\alpha G^{jb}_\beta G^{kc}_\gamma $ 
                     & $(\sigma_\alpha \times\sigma_\beta ) \cdot \sigma_\gamma (\tau_\alpha \times \tau_\beta) \cdot \tau_\gamma$ & 1 \\[.1cm]
\cline{1-6}
$S=1$ &   &   & $  N_c^{-2} \ \epsilon^{ijk} G_\alpha^{ia} G_\beta^{ja}  $ 
                     & $(\sigma_\alpha \times \sigma_\beta) \ \tau_\alpha \cdot \tau_\beta $ & 3 \\
             &   &   & $ N_c^{-3} \ \epsilon^{abc} G_\alpha^{ia} G_\beta^{ib} G_\gamma^{kc} $  
                     & $ (\sigma_\alpha \cdot  \sigma_\beta)  \sigma_\gamma \ (\tau_\alpha \times \tau_\beta) \cdot \tau_\gamma $ & 3 \\[.1cm]
\cline{1-6}
$S=2$ &   &   & $ N_c^{-2} \ [G_\alpha^{ia} G_\beta^{ja}]_{_2} $ & $ [\sigma_\alpha \sigma_\beta]_{_2}   \ \tau_\alpha \cdot \tau_\beta $ & 3  \\
             &   &   & $ \  \ \  N_c^{-3} \ \epsilon^{abc} \Big[ (G_\alpha^{ia} G_\beta^{jb}\epsilon^{ijl})  G_\gamma^{kc}\Big]_{_2} \ \ \  $  
                     & $\ \  \  \Big[ (\sigma_\alpha \times \sigma_\beta ) \sigma_\gamma   \Big]_{_2}  (\tau_\alpha \times \tau_\beta) \cdot \tau_\gamma \ \ \ $ & 2 \\[.1cm]
\cline{1-6}
$S=3$ &   &   & $ N_c^{-3} \ \epsilon^{abc} \Big[ G_\alpha^{ia} G_\beta^{jb} G_\gamma^{kc}\Big]_{_3} $  
                    & $\Big[ \sigma_\alpha \sigma_\beta \sigma_\gamma \Big]_{_3}  (\tau_\alpha \times \tau_\beta) \cdot \tau_\gamma $ & 1\\[.1cm]
\hline
&&&&&17 \\
\hline\hline
\end{tabular}
\caption{Leading-order quark operators and their projection on nucleon
spin-isospin structures. Structures are listed according to their spin content within the nucleonic space. $\alpha, \beta, \gamma$ are a permutation of $123$,
designating on which nucleon the spin and isospin operators act. The multiplicity indicates how
many independent structures are generated by these permutations. The 17 leading-order structures are all parity even and time-reversal even.}
\label{table-leadingops} 
\end{table}

The last column of Table~\ref{table-leadingops} shows the multiplicity of each structure,
obtained by running $\alpha,\beta,\gamma$ over all the permutations of $1,2,3$. 
For the spin-2 structures a
non-trivial constraint reduces the multiplicity of the
$ \Big[(\sigma_\alpha \times \sigma_\beta) \sigma_\gamma\Big]_{_2} 
= \Big\{\Big[(\sigma_1 \times \sigma_2) \sigma_3\Big]_{_2},
   \Big[(\sigma_1 \times \sigma_3) \sigma_2\Big]_{_2},
   \Big[(\sigma_2 \times \sigma_3) \sigma_1\Big]_{_2}  \Big\} $
operator structure from three to two, because Eq.~(\ref{eq:triple}), projected onto 
a symmetric and traceless rank-two tensor, gives:
\begin{eqnarray}
[(A \times B) C]_{_2} + [(B \times C) A]_{_2}  + [(C \times A) B]_{_2}  = 0 \ .
\label{eq:triple2}
\end{eqnarray} 
There are
17 independent structures at leading order. They are all time-reversal even. Further details are given below,
with the leading potential exhibited in Eqs.~(\ref{LOL0}), (\ref{LOL1}), (\ref{LOL2}), and (\ref{LOL3}).

\subsection{Counting all the spin-flavor structures}

However, in order to enumerate all sub-leading structures we find it important to first generalize our counting 
of spin-flavor structures using $SU(4)$ irreps from the NN to the NNN case. In this way we determine the number
of spin-flavor structures we expect to find once we consider all orders in $N_c$.

The number of NNN states is given by
 $\mathbf{4} \otimes \mathbf{4} \otimes \mathbf{4} = 
\mathbf{4} \oplus  \mathbf{20'} \oplus\mathbf{20'} \oplus \mathbf{20}$:
\begin{eqnarray}
\stackrel{\mathbf{4}}{\raisebox{-0.1cm}{\yng(1)}} \otimes \stackrel{\mathbf{4}}{\raisebox{-0.1cm}{\yng(1)}} \otimes 
\stackrel{\mathbf{4}}{\raisebox{-0.1cm}{\yng(1)}} 
&=& 
\stackrel{\mathbf{4}}{\raisebox{-0.8cm}{\yng(1,1,1)}} \oplus 
\stackrel{\mathbf{20'}}{\raisebox{-0.5cm}{\yng(2,1)}} \oplus 
\stackrel{\mathbf{20'}}{\raisebox{-0.5cm}{\yng(2,1)}} \oplus
\stackrel{\mathbf{20}}{\raisebox{-0.1cm}{\yng(3)}} \ . 
\end{eqnarray}
Decomposing these $SU(4)$ irreps into $SU(2)_I \times SU(2)_S \subset SU(4)$ we have 
(as above, the $SU(2)_I \times SU(2)_S$ irreps are labeled by $(2 I +
1, 2 S + 1)$ and only $SU(4)$ irreps are in boldface):
\begin{eqnarray}
\stackrel{\mathbf{20}}{\raisebox{-0.1cm}{\yng(3)}} &=& 
\left(\stackrel{4}{ \raisebox{-0.1cm}{\yng(3)}} \ , \ \stackrel{4}{\raisebox{-0.1cm}{\yng(3)}} \right)  \oplus  
\left( \stackrel{2}{ \raisebox{-0.5cm}{\yng(2,1)}} \ , \
\stackrel{2}{\raisebox{-0.5cm}{\yng(2,1)}} \right) \ , \\
\stackrel{\mathbf{20'}}{\raisebox{-0.5cm}{\yng(2,1)}} &=& 
\left(\stackrel{4}{ \raisebox{-0.1cm}{\yng(3)}} \ , \ \stackrel{2}{\raisebox{-0.5cm}{\yng(2,1)}} \right)  \oplus  
\left(\stackrel{2}{ \raisebox{-0.5cm}{\yng(2,1)}} \ , \ \stackrel{4}{\raisebox{-0.1cm}{\yng(3)}} \right)  \oplus  
\left( \stackrel{2}{ \raisebox{-0.5cm}{\yng(2,1)}} \ , \
\stackrel{2}{\raisebox{-0.5cm}{\yng(2,1)}} \right) \ ,  \\
\stackrel{\mathbf{4}}{\raisebox{-0.8cm}{\yng(1,1,1)}} &=& 
\left(\stackrel{2}{ \raisebox{-0.5cm}{\yng(2,1)}} \ , \
\stackrel{2}{\raisebox{-0.5cm}{\yng(2,1)}} \right)  \ . 
\end{eqnarray}

The number of independent operators $O_{IS}$ of isospin $I$ and spin $S$
that can contribute to $\langle {\rm NNN} | O_{IS} | {\rm NNN}  \rangle $ can now be obtained
by considering the decomposition of $\overline{\mathbf{R}} \otimes \mathbf{R'}$, with 
$\mathbf{R},\mathbf{R'}=\mathbf{4},\mathbf{20'},\mathbf{20}$ into irreps
of $SU(2)_I \times SU(2)_S$. Notice that $\mathbf{20'}$ has to be considered twice.
We are interested in $I=0$ spin-flavor structures, and for those we find:
\begin{eqnarray}
\sum_{R,R'} \overline{\mathbf{R}} \otimes \mathbf{R'} &\supset& 25 (0,0) \oplus 45 (0,3) \oplus 25 (0,5) \oplus 5 (0,7) + \dots
\end{eqnarray}
So, there are 25 independent isoscalar structures of $S=0$, 45 of $S=1$, 25 of
$S=2$ and 5 of $S=3$. There are thus 100 spin-flavor
structures in total, 50 T-even and 50 T-odd. This provides an important check for the explicit construction of
operators that we will describe in the next subsection. The results of that construction
are shown in
Tables~\ref{table-3NL0}--\ref{table-3NL3}, and we indeed find a total of 100 structures.

\subsection{Explicit construction of the three-nucleon operators}

In the following subsubsections we write down, successively, the 3N
potential-energy operators which are built out of  $S=0,1,2,3$  spin-flavor
structures. Since we seek rotational scalars, each spin-flavor structure is
coupled to a momentum structure of equal rank. We therefore use the terms
``$L=a$" and ``$S=a$" interchangeably when referring to the operators that
appear in $V$.  We present explicit expressions up to ${\cal O}(1/N_c)$. 

\subsubsection{$L=S=0$}

A complete set of spin-flavor structures in the $S=0$ sector is given by the $S^{(r)}_\xi$ listed in Table~\ref{table-3NL0}. 
\begin{table}
\begin{tabular}{lclc}
\hline \hline
Operator & Order & T & \qquad Multiplicity \qquad \qquad  \\
\hline
$S^{(0)}_1 = \mathds{1} $ & \qquad \qquad \  $1$ \qquad \qquad \qquad  &  + \qquad & 1 \\
$S^{(0)}_{2-4} = \sigma_\alpha \cdot \sigma_\beta \  \tau_\alpha \cdot \tau_\beta $ & $1$ & + & 3\\
$S^{(0)}_5 = (\sigma_\alpha \times \sigma_\beta) \cdot \sigma_\gamma \ (\tau_\alpha \times \tau_\beta) \cdot \tau_\gamma $ \qquad \qquad & $1$ & $+$ & 1 \\
\hline
$S^{(1)}_{1-3} = \sigma_\alpha \cdot \sigma_\beta  \ (\tau_\alpha \times \tau_\beta) \cdot \tau_\gamma $ & $1/N_c$ & $-$ & 3  \\
$S^{(1)}_{4-6} = (\sigma_\alpha \times \sigma_\beta) \cdot \sigma_\gamma  \  \tau_\alpha \cdot \tau_\beta $ & $1/N_c$ & $-$ & 3 \\
\hline
$S^{(2)}_{1-3} = \tau_\alpha \cdot \tau_\beta $ & $1/N_c^2$ & + & 3 \\
$S^{(2)}_{4-6} = \sigma_\alpha \cdot \sigma_\beta  $ & $1/N_c^2$ & + & 3 \\
$S^{(2)}_{7-12} = \sigma_\alpha \cdot \sigma_\beta  \ \tau_\beta \cdot \tau_\gamma  $ & $1/N_c^2$ & + & 6 \\
\hline
$S^{(3)}_1 = (\tau_\alpha \times \tau_\beta) \cdot \tau_\gamma $ & $1/N_c^3$ & $-$ & 1 \\
$S^{(3)}_2 = (\sigma_\alpha \times \sigma_\beta) \cdot \sigma_\gamma $ & $1/N_c^3$ & $-$ & 1 \\
\hline
 & & & 25 \\
\hline \hline
\end{tabular}
\caption{$S=0$ spin-isospin structures. The order given in the second column is relative to $N_c$. The third column 
indicates the behaviour of each structure structure under time reversal, namely, even (+) or  odd (--), and its  multiplicity 
is given in the last column. Here $\alpha \neq \beta \neq \gamma$ label the nucleon on which each of the spin and isospin operators act. In the 
last line we give the total number of independent structures, obtained as the sum of $M_0^{(0)}=5, M_0^{(1)}=6, M_0^{(2)}=12$
and $M_0^{(3)}=2$.}
\label{table-3NL0}
\end{table}
The superscript $(r)$ indicates the {\em relative order} in $N_c$ at which the
spin-flavor structure appears for the first time (i.e. its lowest order). This corresponds to $r=s+t$ in the Hartree
Hamiltonian, Eq.~(\ref{eq:Hartree}) and essentially counts the number of subleading
operators $S$ and $I$ that contribute to the structure.  The resulting contribution to the 3N
force is obtained after taking into account the overall factor of $N_c$ in
Eq.~(\ref{eq:Hartree}) and  the momentum structure that combines with each spin-flavor
structure to give a rotational scalar, time-reversal-even and parity-even Hamiltonian. 
Each occurrence of a time-reversal-odd momentum ${\bf p}_+,{\bf q}_+$ 
costs an additional power of
$1/N_c$. 

The time-reversal-even spin-flavor structures at order $N_c^0$ and order $1/N_c^2$ can be
straightforwardly incorporated into the potential.  They only need to be multiplied by
arbitrary scalar functions of the vectors ${\bf p}_-$ and ${\bf
q}_-$. We denote the functions, 
which are all of  ${\cal
O} (N_c^0)$, $V_X^m({\bf p}_-^2,
{\bf q}_-^2, {\bf p}_- \cdot {\bf q}_-)$, 
where $X$ runs over the different spin-flavor
structures and $m$ enumerates functions $V$ corresponding to different momentum structures.
Beyond the statement that they are ${\cal O}(N_c^0)$, the
large-$N_c$ expansion sheds no light on the behavior of these functions.

With this notation the ${\cal O} (N_c)$ (leading-order) potential is:
\begin{eqnarray} \label{LOL0}
V_{ L=0}^{N_c}= N_c \sum_{\xi=1}^{M_{0}^{(0)}} V^1_{S_\xi} ({\bf p}_-^2,{\bf q}_-^2,{\bf
p}_- \cdot {\bf q}_-) S_\xi^{(0)} \ , 
\end{eqnarray}
with $M_0^{(0)}=5$ the number of  independent, leading-order, $S=0$ spin-flavor structures
(see Table~\ref{table-3NL0} or Table~\ref{table-leadingops}).  In fact, their contribution once the spatial part of the 3N
state is taken into account is not completely independent, since the functions
$V^{1}_{S_2}$,$V^{1}_{S_3}$ and  $V^{1}_{S_4}$ are related to one another by permutation
symmetry, i.e. the requirement that the total force be symmetric under permutations of all
particle labels. This constraint in the 3N case is, however, more complicated than in the
NN case, and there seems to be no obvious simplification due to permutation symmetry. 

There are thus five spin-flavor  structures that contribute at leading order in $N_c$ to
the $L=0$ part of the 3N potential: the identity, a $ \sigma_\alpha \cdot \sigma_\beta \
\tau_\alpha \cdot \tau_\beta $ structure, where one of the three nucleons is not involved,
and the structure $(\sigma_\alpha \times \sigma_\beta) \cdot \sigma_\gamma \ (\tau_\alpha
\times \tau_\beta) \cdot \tau_\gamma$. Of these, the first two already occur in the NN
potential, and as already discussed above, without knowledge of the ${\bf q}_-$ dependence in
$V$,  we cannot separate their appearance here from the fact that they
contribute to the energy of the NN pairs in the 3N system. 

We now turn our attention to sub-leading corrections to the $L=0$ 3N force. It at first
appears that there are spin-flavor structures which generate  contributions of relative
order $1/N_c$. But in fact the resulting structures are all time-reversal odd. In
consequence they must be multiplied by a time-reversal-odd dot product in order to appear
in the $L=0$ component of the 3N potential. In contrast to the NN case such dot products
exist in this system, e.g. ${\bf p}_+ \cdot {\bf q}_-$. But all of the T-odd ones involve
either ${\bf p}_+$ or ${\bf q}_+$. Thus the first sub-leading contribution is suppressed
by two powers of $1/N_c^2$ relative to leading: one because of the matrix elements of the
spin-flavor structures which appear, and one because of the necessity for a $1/M_N$ factor
in order to generate some non-locality and introduce ${\bf p}_+$ or ${\bf q}_+$. 

At relative order $1/N_c^2$ we also have the 12 structures $S_\xi^{(2)}$ shown in
Table~\ref{table-3NL0}. In addition, 
 the leading structures $S_\xi^{(0)}$ can reappear, now multiplied by two of the $1/N_c$
suppressed dot products, or by one $1/N_c^2$ suppressed dot product of momenta. Using the energy-conservation and
on-shell condition, Eq.~(\ref{eq:onshell}), at  ${\cal O}(1/N_c)$ we find three momentum
structures of ${\cal O}(1/N_c)$ and three structures of
${\cal O}(1/N_c^2)$, all of which involve two momenta. With four momenta there are six new structures
of order ${\cal O}(1/N_c^2)$.  

Lastly, we observe that operators from the LO potential occur again at this order, as they can arise via the replacement of one $G^{ia}/N_c$ by $S^i I^a/N_c^2$ in the Hartree Hamiltonian, as already discussed for the NN potential. We denote this contribution by  $\delta^{(2)} V_{L=0}^{N_c} $ which
stands for 
\begin{eqnarray}\label{LOexpL0} 
\delta^{(2)} V_{ L=0}^{N_c}= N_c^{-1}
\sum_{\xi=1}^{M_{0}^{(0)}} V^1_{S_\xi,1/N_c^2} ({\bf p}_-^2,{\bf q}_-^2,{\bf p}_- \cdot
{\bf q}_-) S_\xi^{(0)} 
\end{eqnarray} 
where the explicit $N_c$ factors ensure that the
$V^1_{S_\xi,1/N_c^2} ({\bf p}_-^2,{\bf q}_-^2,{\bf p}_- \cdot {\bf q}_-)$ are of order
${\cal O}(N_c^0)$. Eqs.~(\ref{LOL0}) and (\ref{LOexpL0}) can be combined, with the 
effect that the functions $V_{S_\xi}^1$ each have their own expansion in $1/N_c^2$.

The
full ${\cal O}(1/N_c)$ piece of the $L=0$ 3N potential is then:
\begin{eqnarray}
V_{L=0}^{1/N_c}&=& \delta^{(2)} V_{L=0}^{N_c} + N_c \sum_{\xi=1}^{M_0^{(2)}} V^{2}_{S_\xi} S_\xi^{(2)} \nonumber\\
&&+ N_c \sum_{\xi=1}^{M_0^{(1)}} \left(  V^{3}_{S_\xi}        {\bf p}_+   \cdot {\bf p}_- 
  +                         V^{4,5}_{S_\xi} {\bf p}_\pm \cdot {\bf q}_\mp  \right)  S_\xi^{(1)} \nonumber\\
&&+ N_c \sum_{\xi=1}^{M_0^{(0)}} \left\{ \ V^{6}_{S_\xi} {\bf p}_+^2 
  +                               V^{7}_{S_\xi} {\bf p}_+ \cdot {\bf q}_+ 
  +                               V^{8}_{S_\xi} {\bf q}_+^2  \right.    \nonumber \\
&&\qquad \ \qquad +  \left.          V^{9}_{S_\xi} ({\bf p}_+ \cdot {\bf p}_-)^2   
  +                               V^{10,11}_{S_\xi} ({\bf p}_\pm \cdot {\bf q}_\mp)^2 \right. \nonumber \\
&&\qquad \ \qquad +         \          V^{12,13}_{S_\xi} ({\bf p}_+ \cdot {\bf p}_-)({\bf p}_\pm \cdot {\bf q}_\mp)  
  +                      \left.   V^{14}_{S_\xi} ({\bf p}_+ \cdot {\bf q}_-)({\bf q}_+ \cdot {\bf p}_-)  \ \right\}  S_\xi^{(0)} \ . 
\label{NLOL0}
\end{eqnarray}
Here $M_0^{(2)}=12$ and $M_0^{(1)}=6$ are obtained summing over the multiplicities shown
in Table~\ref{table-3NL0}. Note that here any momentum structure that involves more powers
of the ${\cal O}(N_c^0)$ momenta ${\bf p}_-$, ${\bf q}_-$, is absorbed in the
${\cal O}(N_c^0)$ scalar functions $V^{2-14}_{S_\xi} ({\bf p}_-^2,{\bf q}_-^2,{\bf p}_-
\cdot {\bf q}_-)$.  

There is no correction at order $1/N_c^3$, due to the T-odd nature of the operators
$S^{(3)}_\xi$ listed above, and the restrictions of parity and time-reversal invariance
regarding the vector dot products which can be considered.  The correction of order
$1/N_c^4$ can be constructed in analogy to the results for $1/N_c^2$ given in
Eq.~(\ref{NLOL0}). For example, the terms involving the $M_0^{(3)}=2$ operators
$S_\xi^{(3)}$ are
\begin{equation}
N_c \sum_{\xi=1}^{M_0^{(3)}} \left(  V^{15}_{S_\xi}        {\bf p}_+   \cdot {\bf p}_- 
  +                         V^{16,17}_{S_\xi} {\bf p}_\pm \cdot {\bf q}_\mp  \right)
S_\xi^{(3)} \ . \\
\end{equation} 
Meanwhile there are many other terms at this order involving $\delta^{(4)} V_{L=0}^{N_c},
\delta^{(2)} V_{L=0}^{1/N_c}   $ and various combinations of dot products of momenta  and the operators
$S_\xi^{(2)}$, $S_\xi^{(1)}$ and $S_\xi^{(0)}$. These are too numerous to list here, but
it is clear that the 
expansion of the potential is in $1/N_c^2$, as was also the case for the NN potential.

Before proceeding to the construction of the $L=1,2,3$ components of the potential it is 
important to explain why there are no cross products in the momentum structures of the
$L=0$ potential.  Triple products like $({\bf p}_-\times{\bf p}_+)\cdot{\bf q}_-$ are
parity odd and should appear together with another triple product giving a structure of
six momenta, which can be written in terms of scalar products alone. Quadruple products
like $({\bf p}_-\times{\bf p}_+)\cdot({\bf p}_-\times{\bf p}_+)$ can also be reduced to
the ones already present, as they involve two contracted epsilon tensors. The identities 
used for this purpose can be found in Appendix~\ref{ap-reduction}, see, specifically, Eq.~(\ref{eq:epseps}).

In summary, for $L=0$ there are five operators at LO, given in Eq.~(\ref{LOL0}), which are built out of T-even spin-isospin structures.
At sub-leading order there are 57 additional operators which are built out of T-even spin-isospin structures,
as well as 18 operators involving T-odd structures. These are listed 
 in  Eq.~(\ref{NLOL0}). Up to ${\cal
O}(1/N_c)$ we presented the explicit expressions for this total of 80 operators, out of
which 17 only depend on ${\bf p}_-, {\bf q}_-$ and correspond  to  a local
potential. 
 
\begin{table}
\begin{tabular}{lclc}
\hline \hline
Operator & Order & T & \qquad Multiplicity \qquad \qquad  \\
\hline
$P^{(0)}_{1-3} = (\sigma_\alpha \times \sigma_\beta) \  \tau_\alpha \cdot \tau_\beta $ & $1$ & + & 3 \\
$P^{(0)}_{4-6} = (\sigma_\alpha \cdot \sigma_\beta) \sigma_\gamma \ (\tau_\alpha \times \tau_\beta) \cdot \tau_\gamma $ \qquad \qquad & $1$ & $+$ & 3 \\
\hline
$P^{(1)}_{1-3} =  \sigma_\alpha $ & \qquad \qquad \  $1/N_c$ \qquad \qquad \qquad  & $-$ \qquad & 3 \\
$P^{(1)}_{4-9} = \sigma_\alpha \ \tau_\alpha \cdot \tau_\beta $ & $1/N_c$ & $-$ & 6 \\
$P^{(1)}_{10-12} = (\sigma_\alpha \times \sigma_\beta) \ (\tau_\alpha \times \tau_\beta) \cdot \tau_\gamma $ & $1/N_c$ & $-$ & 3  \\
$P^{(1)}_{13-15} = (\sigma_\alpha \cdot \sigma_\beta) \sigma_\gamma \ \tau_\alpha \cdot \tau_\beta $ \qquad \qquad & $1/N_c$ & $-$ & 3 \\
$P^{(1)}_{16-21} = (\sigma_\alpha \cdot \sigma_\beta) \sigma_\gamma \ \tau_\beta \cdot \tau_\gamma $ \qquad \qquad & $1/N_c$ & $-$ & 6 \\
\hline
$P^{(2)}_{1-3} = \sigma_\alpha \ (\tau_\alpha \times \tau_\beta) \cdot \tau_\gamma $ & $1/N_c^2$ & + & 3 \\
$P^{(2)}_{4-6} = (\sigma_\alpha \times \sigma_\beta)  $ & $1/N_c^2$ & + & 3 \\
$P^{(2)}_{7-12} = (\sigma_\alpha \times \sigma_\beta)  \ \tau_\beta \cdot \tau_\gamma  $ & $1/N_c^2$ & + & 6\\
\hline
$P^{(3)}_{1-3} = \sigma_\alpha \ \tau_\beta \cdot \tau_\gamma $ & $1/N_c^3$ & $-$ & 3  \\
$P^{(3)}_{4-6} = (\sigma_\alpha \cdot \sigma_\beta) \sigma_\gamma $ \qquad \qquad & $1/N_c^3$ & $-$ & 3 \\
\hline
& & & 45 \\
\hline \hline
\end{tabular}
\caption{$S=1$ spin-isospin structures, as in Table~\ref{table-3NL0}. In the 
last line we give the total number of independent structures, obtained as the sum of $M_1^{(0)}=6, M_1^{(1)}=21, M_1^{(2)}=12$
and $M_1^{(3)}=6$.}
\label{table-3NL1}
\end{table}

\begin{table}
\begin{tabular}{lclc}
\hline \hline
Operator & Order & T & \qquad Multiplicity \qquad \qquad  \\
\hline
$D^{(0)}_{1-3} = [\sigma_\alpha \sigma_\beta]_{_2} \  \tau_\alpha \cdot \tau_\beta $ & $1$ & + & 3 \\
$D^{(0)}_{4,5} = \Big[(\sigma_\alpha \times \sigma_\beta) \sigma_\gamma\Big]_{_2} \ (\tau_\alpha \times \tau_\beta) \cdot \tau_\gamma $ \qquad \qquad & $1$ & $+$ & 2 \\[.1cm]
\hline
$D^{(1)}_{1-3} = [\sigma_\alpha \sigma_\beta]_{_2} \ (\tau_\alpha \times \tau_\beta) \cdot \tau_\gamma $ & $1/N_c $ & $-$ & 3 \\
$D^{(1)}_{4,5} = \Big[(\sigma_\alpha \times \sigma_\beta) \sigma_\gamma\Big]_{_2} \  \tau_\alpha \cdot \tau_\beta $ \qquad \qquad & $1/N_c$ & $-$ & 2\\
$D^{(1)}_{6-9} = \Big[(\sigma_\alpha \times \sigma_\beta) \sigma_\gamma\Big]_{_2} \  \tau_\beta \cdot \tau_\gamma $ \qquad \qquad & $1/N_c$ & $-$ & 4\\[.1cm]
\hline
$D^{(2)}_{1-3} = [\sigma_\alpha \sigma_\beta]_{_2} $ & \qquad \qquad $1/N_c^2$\qquad \qquad \qquad  & + & 3 \\
$D^{(2)}_{4-9} = [\sigma_\alpha \sigma_\beta]_{_2} \  \tau_\beta \cdot \tau_\gamma $ & $1/N_c^2$ & + & 6 \\
\hline
$D^{(3)}_{1,2} = \Big[(\sigma_\alpha \times \sigma_\beta) \sigma_\gamma\Big]_{_2} $ \qquad \qquad & $1/N_c^3$ & $-$ & 2\\[.1cm]
\hline
& & & 25 \\
\hline
\hline
\end{tabular}
\caption{$S=2$ spin-isospin structures, as in Table~\ref{table-3NL0}. In the 
last line we give the total number of independent structures, obtained as the sum of $M_2^{(0)}=5, M_2^{(1)}=9, M_2^{(2)}=9$
and $M_2^{(3)}=2$.}
\label{table-3NL2}
\end{table}

\begin{table}
\begin{tabular}{lclc}
\hline \hline
Operator & Order & T & \qquad Multiplicity \qquad \qquad  \\
\hline
$F^{(0)}_1 =[\sigma_\alpha \sigma_\beta \sigma_\gamma]_{_3} \ (\tau_\alpha \times \tau_\beta) \cdot \tau_\gamma 
$ \qquad \qquad & $1$ & $+$ & 1\\
\hline
$F^{(1)}_{1-3} =[\sigma_\alpha \sigma_\beta \sigma_\gamma]_{_3} \  \tau_\alpha \cdot \tau_\beta 
$ & \qquad \qquad \  $1/N_c$ \qquad \qquad \qquad& $-$ & 3 \\
\hline
$F^{(3)}_1 =[\sigma_\alpha \sigma_\beta \sigma_\gamma]_{_3} 
$ & $1/N_c^3$ & $-$ & 1 \\
\hline
& & & 5 \\
\hline \hline 
\end{tabular}
\caption{$S=3$ spin-isospin structures, as in Table~\ref{table-3NL0}. In the 
last line we give the total number of independent structures, obtained as the sum of $M_3^{(0)}=1, M_3^{(1)}=3, M_3^{(2)}=0$
and $M_3^{(3)}=1$.}
\label{table-3NL3}
\end{table}

\subsubsection{$L=S=1$}

In contrast to the two-nucleon potential, the three-nucleon potential contains $S=L=1$
terms at leading order.  That is because vector spin-flavor structures can be constructed out of two
and three $G$'s, see Table~\ref{table-leadingops}, and then contracted with the P-even,
T-even  cross product ${\bf p}_- \times {\bf q}_-$. No such time-reversal-even cross
product exists in the NN system at leading order.  The
leading-order $L=1$ force is then: 
\begin{equation}
V_{L=1}^{N_c}= N_c \sum_{\xi=1}^{M_1^{(0)}} V_{P_\xi}^{1}({\bf p}_- \times {\bf q}_-) \cdot P_\xi^{(0)}.
\label{LOL1}
\end{equation}
Here $M_1^{(0)}=6$ is the number of leading-order $S=1$ structures, see
Table~\ref{table-3NL1}. Since ${\bf q}_-$ appears in the momentum structure of
Eq.~(\ref{LOL1}) they cannot occur in the NN force, and are unambiguously the result of
3N interactions. However, we note that, once again, the functions $V^1_{P_{1-3}}$ will be related to one another  through permutation symmetry, as will the functions $V^1_{P_{4-6}}$.  

Operators with matrix elements suppressed by $1/N_c$ are easily obtained, see
Table~\ref{table-3NL1} where they are listed as $P_\xi^{(1)}$, with $\xi=1,\dots ,
M_1^{(1)}$, and $M_1^{(1)}=21$. However, as is displayed in the table, these are all
time-reversal odd. Thus they must be contracted with T-odd cross products, and this costs
another power of $1/N_c$, since it mandates that ${\bf p}_+$ or ${\bf q}_+$ be involved.
Thus the order of such contributions is $1/N_c^2$ relative to leading. At this order the
T-even structures $P_\xi^{(2)}$ also appear, contracted with ${\bf p}_- \times {\bf q}_-$,
the leading-order momentum structure.  The $L=1$ 3N force of order $1/N_c^2$ is therefore
of the form: 
\begin{eqnarray} \label{NLOL1}
V_{L=1}^{1/N_c}&=& \delta^{(2)} V_{L=1}^{N_c} + 
N_c \sum_{\xi=1}^{M_1^{(2)}} V_{P_\xi}^{2}({\bf p}_- \times {\bf q}_-) \cdot P_\xi^{(2)}
\nonumber\\
&&+ N_c  \sum_{\xi=1}^{M_1^{(1)}} \left\{ \ V_{P_\xi}^{3} {\bf p}_+ \times {\bf p}_- 
  +                        V_{P_\xi}^{4,5} {\bf p}_\pm \times {\bf q}_\mp   
  +                        V_{P_\xi}^{6} {\bf q}_+ \times {\bf q}_-
\right. \nonumber\\
&& \qquad \qquad \ +  \left. \left(  V^{7}_{P_\xi}        {\bf p}_+   \cdot {\bf p}_- 
  +                         V^{8,9}_{P_\xi} {\bf p}_\pm \cdot {\bf q}_\mp  \right) 
                            ({\bf p}_- \times {\bf q}_-) \ \right\}  \cdot P_\xi^{(1)}
\nonumber\\
&&+ N_c \sum_{\xi=1}^{M_1^{(0)}} \left\{ \ V_{P_\xi}^{10} ({\bf p}_+ \times {\bf q}_+) \right.  \nonumber\\
&&\qquad \qquad \ + \left.  \left( V_{P_\xi}^{11} {\bf p}_+^2 
                              + V_{P_\xi}^{12} {\bf p_+ \cdot q_+} 
                              + V_{P_\xi}^{13} {\bf q}_+^2 \right) ({\bf p}_- \times {\bf q}_-) \right.  \nonumber\\
&& \qquad \qquad \ +  \left. \left(  V^{14}_{P_\xi}        {\bf p}_+   \cdot {\bf p}_- 
  +                         V^{15,16}_{P_\xi} {\bf p}_\pm \cdot {\bf q}_\mp  \right) 
                            ({\bf p}_+ \times {\bf p}_-) \ \right. 
\nonumber\\
&& \qquad \qquad \ +  \left. \left(  V^{17}_{P_\xi}        {\bf p}_+   \cdot {\bf p}_- 
  +                         V^{18,19}_{P_\xi} {\bf p}_\pm \cdot {\bf q}_\mp  \right) 
                            ({\bf p}_+ \times {\bf q}_-) \ \right. 
\nonumber\\
&& \qquad \qquad \ +  \left. \left(  V^{20}_{P_\xi}        {\bf p}_+   \cdot {\bf p}_- 
  +                         V^{21,22}_{P_\xi} {\bf p}_\pm \cdot {\bf q}_\mp  \right) 
                            ({\bf p}_- \times {\bf q}_+) \ \right. 
\nonumber\\
&& \qquad \qquad \ +  \left. \left(  V^{23}_{P_\xi}        {\bf p}_+   \cdot {\bf p}_- 
  +                         V^{24,25}_{P_\xi} {\bf p}_\pm \cdot {\bf q}_\mp  \right) 
                            ({\bf q}_+ \times {\bf q}_-) 
                   \ \right\} \cdot P_\xi^{(0)} \nonumber\\
\end{eqnarray}
In this equation $M_1^{(2)}=12$ and $M_1^{(1)}=21$.  Using two momenta there is only one
structure of ${\cal O}(1)$. There are four structures of ${\cal O}(1/N_c)$  and one
structure of ${\cal O}(1/N_c^2)$. With four momenta  there are three structures at  ${\cal O}(1/N_c)$
and 15 new structures of ${\cal O}(1/N_c^2)$.  Triple products can be
eliminated using the identity (\ref{eq:triple}).

Once again, the new spin-flavor structures that appear
at ${\cal O}(1/N_c^3)$ are all T-odd. Thus they must be combined with ${\bf p}_- \times {\bf
q}_+$, or one of the three other T-odd cross products involving a ${\bf +}$ vector, to
yield something appropriate for inclusion in $V_{L=1}$. The overall result is then a
contribution to the 3N force of relative order $1/N_c^4$. The expansion for $V_{L=1}$ is, like that
for $V_{L=0}$, an expansion in $1/N_c^2$. 

In summary, for $L=1$ there are six operators at LO, given in Eq.~(\ref{LOL1}), which are built out 
of T-even spin-isospin structures. At sub-leading order there 108 new operators involving T-even spin-isospin structures, and 
147 operators involving T-odd ones, see  Eq.~(\ref{NLOL1}). Up to ${\cal
O}(1/N_c)$ we presented the explicit expressions for these 261 operators, out of
which 18 involve only ${\bf p}_-, {\bf q}_-$, and so correspond  to  a local
potential. 

\subsubsection{$L=S=2$}

The leading 3N spin-flavor structures with $S=2$ are constructed out of $G$'s and $\mathds{1}$'s as
shown in Table~\ref{table-leadingops}.  A Cartesian rank-two tensor with $S=2$, constructed out of
two vector quantities $A^i$ and $B^j$, is symmetric and traceless in its two indices and
will be denoted $ [ A_i B_j ]_{_2}$ (see Eq.~(\ref{eq:rank2tensor})).  Subleading
structures are obtained after introducing a growing number of $S$ and $I$ operators,
following Eq.~(\ref{eq:Hartree}). 
 
The complete set of resulting spin-flavor structures is displayed in
Table~\ref{table-3NL2}. This time we have $M_2^{(0)}=5$. The five LO structures must be
contracted with $L=2$ tensors constructed out of ${\bf p}_-$ and ${\bf q}_-$ to obtain the
LO contribution to the 3N potential. 

We reiterate that Eq.~(\ref{eq:triple2}) explains why there are only two structures  $D^{(0)}_{4,5}$,
instead of the multiplicity three in similar structures with three spin operators in the $L=1$ case.
Similar reductions in multiplicity occur for sub-leading $S=2$ spin-flavor structures too.

The leading force is then:
\begin{eqnarray} \label{LOL2}
V_{L=2}^{N_c}= N_c \sum_{\xi=1}^{M^{(0)}_2} \left\{ \  V_{D_\xi}^{1} \ [{\bf p}_- {\bf p}_-]_{_2} 
+ V_{D_\xi}^{2} \ [{\bf p}_- {\bf q}_-]_{_2} 
+ V_{D_\xi}^{3} \ [{\bf q}_- {\bf q}_-]_{_2} \  \right\}  \cdot  D_\xi^{(0)} \ , 
\end{eqnarray}
with the sum over the $M_2^{(0)}=5$ structures as listed in Table~\ref{table-3NL2}. Note that the first three structures
in the sum, i.e. $D_{1-3}^{(0)}$, contracted with the appropriate ${\bf p}_-$, already occur in the NN potential. 
Only ${\bf q}_-$ dependence in $ V_{D_{1-3}}^{1} $ would reveal it is a ``true" 3N force. 

Using analogous arguments to those already discussed for the $L=0$ and $1$ cases, we obtain the subleading
$L=S=2$
contribution to the 3NF:
\begin{eqnarray} \label{NLOL2}
V_{L=2}^{1/N_c}&=& \delta^{(2)} V_{L=2}^{N_c} + N_c \sum_{\xi=1}^{M^{(2)}_2} \left\{ \  
                                            V_{D_\xi}^{4} \ [{\bf p}_- {\bf p}_-]_{_2} 
                                          + V_{D_\xi}^{5} \ [{\bf p}_- {\bf q}_-]_{_2} 
                                          + V_{D_\xi}^{6} \ [{\bf q}_- {\bf q}_-]_{_2}  
                                          \ \right\}  \cdot  D_\xi^{(2)} \nonumber \\
&& + N_c  \sum_{\xi=1}^{M^{(1)}_2} \left\{ \  
                             V_{D_\xi}^{7}  \       [{\bf p}_+ {\bf p}_-]_{_2} 
                           + V_{D_\xi}^{8,9} \ [{\bf p}_\pm {\bf q}_\mp]_{_2} 
                           + V_{D_\xi}^{10}         \ [{\bf q}_+ {\bf q}_-]_{_2}  
                               \right.  \nonumber \\
&& \ \qquad \qquad + \left(  
                        V_{D_\xi}^{11} \ {\bf p}_+ \cdot {\bf p}_- + V_{D_\xi}^{12,13} \ {\bf p_\pm \cdot q_\mp} 
                   \right) [{\bf p}_- {\bf p}_-]_{_2} \nonumber \\ 
&& \ \qquad \qquad + \left(  
                        V_{D_\xi}^{14} \ {\bf p}_+ \cdot {\bf p}_- + V_{D_\xi}^{15,16} \ {\bf p_\pm \cdot q_\mp} \right) 
[{\bf p}_- {\bf q}_-]_{_2} \nonumber \\
&& \ \qquad \qquad + \left. \left(  V_{D_\xi}^{17} \ {\bf p}_+ \cdot {\bf p}_- + V_{D_\xi}^{18,19} \ {\bf p_\pm \cdot q_\mp} \right) 
[{\bf q}_- {\bf q}_-]_{_2} 
\    \right\} \cdot  D_\xi^{(1)} \nonumber \\
&& +  N_c \sum_{\xi=1}^{M^{(0)}_2} \left\{ \  V_{D_\xi}^{20} \ [{\bf p}_+ {\bf p}_+]_{_2} 
                     + V_{D_\xi}^{21} \ [{\bf p}_+ {\bf q}_+]_{_2} 
                     + V_{D_\xi}^{22} \ [{\bf q}_+ {\bf q}_+]_{_2} 
                              \right. \nonumber \\
&&\ \qquad \qquad +  \left( V_{D_\xi}^{23} \ {\bf p}_+^2 + V_{D_\xi}^{24} \ {\bf p_+ \cdot q_+} + V_{D_\xi}^{25} \ {\bf q}_+^2 \right)  [{\bf p}_- {\bf p}_-]_{_2}  \nonumber\\
&&\ \qquad \qquad +  \left( V_{D_\xi}^{26} \ {\bf p}_+^2 + V_{D_\xi}^{27} \ {\bf p_+ \cdot q_+} + V_{D_\xi}^{28} \ {\bf q}_+^2 \right)  [{\bf p}_- {\bf q}_-]_{_2}  \nonumber\\
&&\ \qquad \qquad +  \left( V_{D_\xi}^{29} \ {\bf p}_+^2 + V_{D_\xi}^{30} \ {\bf p_+ \cdot q_+} + V_{D_\xi}^{31} \ {\bf q}_+^2 \right)  [{\bf q}_- {\bf q}_-]_{_2}  \nonumber\\
&&\ \qquad \qquad +  \left( V_{D_\xi}^{32}  \ {\bf p_+ \cdot p_-} + V_{D_\xi}^{33,34} \ {\bf p_\pm \cdot q_\mp} \right)  [{\bf p}_+ {\bf p}_-]_{_2}  \nonumber\\
&&\ \qquad \qquad +  \left( V_{D_\xi}^{35}  \ {\bf p_+ \cdot p_-} + V_{D_\xi}^{36,37} \ {\bf p_\pm \cdot q_\mp}\right) [{\bf p}_+ {\bf q}_-]_{_2}  \nonumber\\
&&\ \qquad \qquad +  \left( V_{D_\xi}^{38}  \ {\bf p_+ \cdot p_-} + V_{D_\xi}^{39,40} \ {\bf p_\pm \cdot q_\mp}\right) [{\bf p}_- {\bf q}_+]_{_2}  \nonumber\\
&&\ \qquad \qquad +  \left. \left( V_{D_\xi}^{41} \  {\bf p_+ \cdot p_-} + V_{D_\xi}^{42,43} \ {\bf p_\pm \cdot q_\mp} \right)  [{\bf q}_+ {\bf q}_-]_{_2} 
    \ \right\} \cdot  D_\xi^{(0)} \ , 
\end{eqnarray}
where $M^{(1)}_2=9,M^{(2)}_2=9$. In writing Eq.~(\ref{NLOL2}) we employed three tensors built out of two momenta at ${\cal O}(1)$, four at ${\cal
O}(1/N_c)$ and three at ${\cal O}(1/N_c^2)$; with four momenta there are 9
structures at ${\cal O}(1/N_c)$ and 21 at ${\cal O}(1/N_c^2)$. Of course, all structures are again multiplied by the usual arbitrary functions of ${\bf p}_-^2$, ${\bf q}_-^2$, and ${\bf p}_- \cdot {\bf q}_-$. As in the $L=S=0$ and $L=S=1$ cases, the next-to-next-to-leading-order contributions to $V$ appear at relative order $1/N_c^4$, as a consequence of parity and time reversal. 

In summary, for this $L=S=2$ part of the 3N force there are 15 operators at LO, which are built out 
of T-even spin-isospin structures. These are given in Eq.~(\ref{LOL2}). There are 264
new operators at relative order $1/N_c^2$, 147 (117) of which are based on T-even (T-odd) spin-flavor structures, see 
Eq.~(\ref{NLOL2}). Up to ${\cal O}(1/N_c)$ we presented the explicit expressions for all of these 279 operators. Amongst these are
42 operators that only depend on ${\bf p}_-, {\bf q}_-$ and could appear in a local
potential. 
 
\subsubsection{$L=S=3$}

These operators have no NN analog. The rank-three Cartesian tensor with $L=3$ 
that can be constructed out of three vectors $A,B,C$  
is symmetric and traceless, namely
\begin{eqnarray}
{[}A_i B_j C_k]_{_3} &=& A_i B_j C_k + A_j B_k C_i + A_k B_i C_j + A_j B_i C_k + A_i B_k C_j + A_k B_j C_i \nn \\
&& - \frac25 \delta_{ij} (A \cdot B \ C_k +  A \cdot C \ B_k + B \cdot C \ A_k) \nn \\ 
&&  - \frac25 \delta_{ik} (A \cdot B \ C_j +  A \cdot C \ B_j + B \cdot C \ A_j) \nn \\
&&  - \frac25 \delta_{jk} (A \cdot B \ C_i +  A \cdot C  \ B_i + B \cdot C \ A_i) \ . 
\end{eqnarray}
There is just one leading-order spin-flavor structure, three suppressed by $1/N_c$ and one
suppressed by $1/N_c^3$, as shown in Table~\ref{table-3NL3}. Note that there are no
$1/N_c^2$ structures in this case, because of isospin conservation, as three $\sigma$
operators would need to appear along with one $\tau$ to generate a structure at that order.
Table~\ref{table-3NL3} again shows that the suppressed structures are T-odd, and so
ultimately lead to contributions to the 3N force that are down by $1/N_c^2$ and
$1/N_c^4$ respectively.

Because of parity conservation, we need at least four momenta to construct the $L=3$
component of the potential.  There are three $L=3$, P-even, T-even, momentum tensors at
${\cal O}(1)$. Both ${\bf q}_-$ and ${\bf p}_-$ are needed to construct these, as, e.g. ${\bf q}_- {\bf q}_{-} {\bf q}_{-} {\bf q}_{-} $ only
contains $L=0,2,4$ components.  The leading operator is
\begin{eqnarray} \label{LOL3}
V_{L=3}^{N_c} =
N_c \left\{ \  
V^{1}_{F_1} [({\bf p}_- \times {\bf q}_-)  {\bf p}_- {\bf p}_-]_{_3} + 
V^{2}_{F_1} [({\bf p}_- \times {\bf q}_-)  {\bf p}_- {\bf q}_-]_{_3} + 
V^{3}_{F_1} [({\bf p}_- \times {\bf q}_-)  {\bf q}_- {\bf q}_-]_{_3} 
\ \right\}
\cdot F^{(0)}_1 \ . \nonumber \\
\end{eqnarray}

The subleading $L=3$ force contains $L=3$, P-even T-odd
tensors involving three ${\bf q}_-$ or ${\bf p}_-$ vectors, together with one ${\bf p}_+$
or ${\bf q}_+$, contracted with the three $F_{1-3}^{(1)}$ structures.  Also appearing at
this order are the terms which involve the leading spin-flavor structure, contracted with
P-even T-even tensors in which two of the four vectors are ${\bf p}_+$ or ${\bf q}_+$. Meanwhile, 
the
subleading correction to the LO structure $F_1^{(0)}$ corresponds to $s=t=1$ in
Eq.~(\ref{eq:Hartree}) and is given by $\delta^{(2)} V_{L=3}^{N_c}$. Thus, finally we obtain
\begin{eqnarray} \label{NLOL3}
V_{L=3}^{1/N_c} &=& \delta^{(2)} V_{L=3}^{N_c}  \nonumber \\
&& + N_c  \sum_{\xi=1}^{M^{(1)}_3}  \left\{ \  
V^{4}_{F_\xi} [({\bf p}_+ \times {\bf p}_-)  {\bf p}_- {\bf p}_-]_{_3} + 
V^{5}_{F_\xi} [({\bf p}_+ \times {\bf p}_-)  {\bf p}_- {\bf q}_-]_{_3} + 
V^{6}_{F_\xi} [({\bf p}_+ \times {\bf p}_-)  {\bf q}_- {\bf q}_-]_{_3} \right. \nonumber \\
&& \quad \qquad   + \left.  
V^{7,8}_{F_\xi} [({\bf p}_\pm \times {\bf q}_\mp)  {\bf p}_- {\bf p}_-]_{_3} + 
V^{9,10}_{F_\xi} [({\bf p}_\pm \times {\bf q}_\mp)  {\bf p}_- {\bf q}_-]_{_3} + 
V^{11,12}_{F_\xi} [({\bf p}_\pm \times {\bf q}_\mp)  {\bf q}_- {\bf q}_-]_{_3}  \right. \nonumber \\
&& \quad \qquad  + \left.  
V^{13}_{F_\xi} [({\bf q}_+ \times {\bf q}_-)  {\bf p}_- {\bf p}_-]_{_3} + 
V^{14}_{F_\xi} [({\bf q}_+ \times {\bf q}_-)  {\bf p}_- {\bf q}_-]_{_3} + 
V^{15}_{F_\xi} [({\bf q}_+ \times {\bf q}_-)  {\bf q}_- {\bf q}_-]_{_3} 
\ \ \  \right\} \cdot F^{(1)}_\xi \nonumber \\
&&+ N_c \left\{ \  V^{16}_{F_1} [({\bf p}_+ \times {\bf p}_-)  {\bf p}_+ {\bf p}_-]_{_3} \right. \nonumber \\
&& \ \qquad \ + \left.
        V^{17,18}_{F_1} [({\bf p}_+ \times {\bf p}_-)  {\bf p}_\pm {\bf q}_\mp]_{_3}  
      + V^{19}_{F_1} [({\bf p}_+ \times {\bf p}_-)  {\bf q}_+ {\bf q}_-]_{_3}  \right.  \nonumber \\
&& \ \qquad \ + \left.
       V^{20,21}_{F_1} [({\bf p}_\pm \times {\bf q}_\mp)  {\bf p}_+ {\bf p}_-]_{_3} 
      +V^{22,23}_{F_1} [({\bf p}_\pm \times {\bf q}_\mp)  {\bf p}_+ {\bf q}_-]_{_3} 
      \right.  \nonumber \\
&& \ \qquad \ + \left.
      V^{24,25}_{F_1} [({\bf p}_\pm \times {\bf q}_\mp)  {\bf q}_+ {\bf p}_-]_{_3} 
      +V^{26,27}_{F_1} [({\bf p}_\pm \times {\bf q}_\mp)  {\bf q}_+ {\bf q}_-]_{_3} 
\right.  \nonumber \\
&& \ \qquad \  + \left.
      V^{28,29}_{F_1} [({\bf q}_+ \times {\bf q}_-)  {\bf p}_\pm {\bf q}_\mp]_{_3} 
      +V^{30}_{F_1} [({\bf q}_+ \times {\bf q}_-)  {\bf q}_+ {\bf q}_-]_{_3}   \ \right\}
\cdot F^{(0)}_1 \ , 
\end{eqnarray}
where $M^{(1)}_3=3$. There are 12 momentum structures at ${\cal O}(1/N_c)$ and 15 momentum
structures at ${\cal O}(1/N_c^2)$.  Here we used again Eq.~(\ref{eq:triple2}) to reduce
the number of momentum structures. The usual arguments  show that the next correction,
which involves the $F_1^{(3)}$ structure, is down by $1/N_c^4$ compared to the leading
contribution, again because of the need to contract $F_1^{(3)}$ with a P-even, T-odd
tensor. 

In summary, for $L=3$ there are three operators at LO built out of T-even
spin-isospin structures, given in Eq.~(\ref{LOL3}), and 51 additional (15
corresponding to new T-even and 36 to T-odd structures) at relative order
$1/N_c^2$, given in  Eq.~(\ref{NLOL3}). We presented the explicit expressions
for these 54 operators which occur up to ${\cal O}(1/N_c)$. Out of these three
depend solely on ${\bf p}_-, {\bf q}_-$ and could be part of a local potential.

\section{Summary and Discussion}

\label{sec-conclusion}

We have classified all the spin-flavor structures that can contribute to the
three-nucleon force (3NF) and power counted these structures in the $1/N_c$
expansion. The leading-order (LO) part of the 3NF is constructed from $G^{ia}/N_c$ and the unit
operator, since these are the quark operators that have nucleon matrix elements that are ${\cal O}(1)$.
Isospin-invariant structures like
\begin{equation}
\mathds{1}_\alpha \mathds{1}_\beta \mathds{1}_\gamma , 
\qquad N_c^{-2} G^{ia}_\alpha \mathds{1}_\beta G^{j a}_\gamma , 
\qquad N_c^{-3} \epsilon^{abc} G^{ia}_\alpha G^{jb}_\beta G^{kc}_\gamma,
\label{eq:LOops}
\end{equation}
with $\alpha$, $\beta$, and $\gamma$ labelling the three nucleons, are the leading
contributions. Contraction of these structures with spatial tensors of the appropriate rank, built from the ${\cal O}(1)$ momenta ${\bf p}_-$ and ${\bf q}_-$, together with a re-expression in
terms of the angular momentum content of these structures, and use of the reduction of $G^{ia}$ to
$\sigma^i \tau^a$ when restricted to the nucleon subspace, produces the LO force:
\begin{eqnarray} \label{theLO}
V_{3{\rm N}}^{N_c} &=&N_c \sum_{\xi=1}^{M_{0}^{(0)}} V^1_{S_\xi} S_\xi^{(0)}
+ N_c \sum_{\xi=1}^{M_1^{(0)}} V_{P_\xi}^{1}({\bf p}_- \times {\bf q}_-) \cdot P_\xi^{(0)} \\
&&  +   N_c \sum_{\xi=1}^{M^{(0)}_2} \left\{ \  V_{D_\xi}^{1} \ [{\bf p}_- {\bf p}_-]_{_2} 
+ V_{D_\xi}^{2} \ [{\bf p}_- {\bf q}_-]_{_2} 
+ V_{D_\xi}^{3} \ [{\bf q}_- {\bf q}_-]_{_2} \  \right\}  \cdot  D_\xi^{(0)}  \nonumber \\ 
&& + 
N_c \left\{ \  
V^{1}_{F_1} [({\bf p}_- \times {\bf q}_-)  {\bf p}_- {\bf p}_-]_{_3} + 
V^{2}_{F_1} [({\bf p}_- \times {\bf q}_-)  {\bf p}_- {\bf q}_-]_{_3} + 
V^{3}_{F_1} [({\bf p}_- \times {\bf q}_-)  {\bf q}_- {\bf q}_-]_{_3} 
\ \right\}
\cdot F^{(0)}_1. \nonumber 
\end{eqnarray}
Here the $S,P,D,F$ spin-flavor structures are given in Tables~\ref{table-3NL0}, \ref{table-3NL1},
\ref{table-3NL2}, \ref{table-3NL3}. The corresponding multiplicities
are $M_{0,1,2,3}^{(0)}=5,6,5,1 $.
Including all the independent  momentum structures, to leading order we have 29 operators
distributed as $5,6,15$ and 3 operators with $L=0,1,2,3$ respectively.  

It follows straightforwardly  that our LO force contains the structures present in the
Fujita-Miyazawa three-nucleon potential, Eq.~(\ref{eq:FM}). Indeed, the only structure
beyond the Fujita-Miyazawa result is the unit operator which is added to ${V}_{ijk}^{2
\pi}$ in most modern implementations of the 3NF. Of course, models of the 3NFs contain
specific predictions for the coefficient functions $V_{L_\xi}^{m}$. The large-$N_c$
expansion can say nothing about these functions beyond the statement that they should 
be ``natural", i.e. ${\cal O}(1)$; the insights from large-$N_c$ reside in the statements
regarding the overall size that different 
spin-isospin-momentum structures within the 3NF should have.

The LO 3NF contains spin-dependent forces, but it does not contain the spin-orbit forces
that have been proposed as a solution to the $A_y$ puzzle (see, e.g. Ref.~\cite{Ki99}).
The $A_y$ puzzle is not straightforwardly resolved by $1/N_c$ power counting arguments. 

Spin-orbit forces, together with several other operators, all of which we have tabulated
in Section~\ref{sec-3N}, appear at ${\cal O}(1/N_c^2)$ compared to leading. We have also
shown that the next-to-next-to-leading correction to the 3NF is at order $1/N_c^4$ relative to LO. The
NNN force is therefore, like the NN force, an expansion in $1/N_c^2$.  We have given
explicit expressions for the 674 operators that appear in the 3N potential up to (overall)
order $1/N_c$ in  Eqs.~(\ref{theLO}),
 (\ref{NLOL0}), (\ref{NLOL1}),
 (\ref{NLOL2}) and (\ref{NLOL3}).

Many of these operators involve non-localities and time-reversal-odd momentum structures. 
For a local 3NF
only time-reversal-even momentum structures involving ${\bf p}_-$ or ${\bf q}_-$ 
can occur. Such structures occur in both the leading and sub-leading 3NF, but do
not occur at higher orders, where the presence of at least one time-reversal-odd momentum is required. 
Taking into account the different momentum
structures which satisfy this constraint, at relative order $1/N_c^2$ we have $12$, $12$ and 27
operators with $L=0,1,2$ respectively. No $L=3$ local operator occurs at this relative order in the expansion.
Combining these operators with the 29 LO operators yields
a total of 80 operators that constitute the most general  basis for a local
3NF. These operators can be easily read off from Eqs.~(\ref{theLO}),
(\ref{NLOL0}), (\ref{NLOL1}),
(\ref{NLOL2}) and (\ref{NLOL3}).
In a recent paper~\cite{Krebs:2013kha} the authors needed a basis of 89 operators
to obtain the most general contribution of a local 3NF. Their operator  basis  is somewhat
different from ours, so a comparison is not immediate.  An important subject for future investigation
is the
relation between the two sets of operators, and a determination of the minimal basis of operators 
for a general, local 3NF.

We have not discussed the constraints imposed by the permutation group on our analysis. In
the NN case such considerations resulted in the elimination of the spin structure
$\sigma_1 - \sigma_2$. In the 3N case such constraints will impose relations between the
different coefficient functions we have used in our expansion.  Since the 3N coefficient
functions depend on three rotational scalars it seems unlikely that a general,
permutation-group-based argument can be used to eliminate an operator structure from the
3NF---at least in the absence of additional assumptions about the coefficient functions
themselves. A permutation-group analysis of the structures we have obtained would be a
useful step towards understanding the particular 3N partial waves which the different
operator structures we have obtained contribute to.  

It would also be interesting to test whether adding the $1/N_c^2$ structures we have
listed here to phenomenological 3NFs improves the description of few-nucleon scattering
data and light-nuclear spectra. Recently developed three-body potentials like the lllinois
force~\cite{Pi01} or that derived from $\chi$PT~\cite{Be08,Be11,Krebs:2013kha} include several of these structures. Matching such 3NFs to
our large-$N_c$ expressions is appreciably more involved than in the NN case analyzed in
Ref.~\cite{KM97}, but nevertheless, they could be matched to the list of
operators presented here. This would illuminate precisely which structures are present in
these particular potentials, and whether the relative size of the different contributions is well predicted by the $1/N_c$ expansion. 

\acknowledgments{This work was supported by the US Department of Energy under grant
DE-FG02-93ER40756. C.S. thanks the Institute of
Nuclear and Particle Physics at Ohio University for its warm hospitality during a long-term visit where most of this work was done. 
We are both grateful to the Institute for Nuclear Theory,
for support during INT program 10-1, ``Simulations and Symmetries: Cold Atoms, QCD, and Few-hadron Systems",
which facilitated the early stages of this project. D.R.P. thanks Steve Pieper and Bob Wiringa for useful discussions. 
C.S. thanks the DESY Theory Group for its hospitality during the 
final stage of this project.}

\appendix

\section{An explicit example of operator reduction}

For completeness, it is worth discussing an explicit example of the operator reduction
Eq.~(\ref{eq:manyG}). Consider the case of the product of two $G$'s, which can be written as
\begin{eqnarray}
G^{ia} G^{jb} &=& \frac12 \{G^{ia},G^{jb}\} + \frac12 {[}G^{ia},G^{jb}] \ .
\end{eqnarray}
The commutator contains the antisymmetric terms in $(ia)$ and $(jb)$ and is suppressed by $1/N_c^2$. The symmetric part in 
$(ia)$ and $(jb)$ can be written as a tensor $W^{(ij),(ab)}$ which is symmetric in the
spatial indices $i,j$ and symmetric in the isospin
indices $a,b$ independently, and a tensor $W^{[ij],[ab]}$ that is antisymmetric in the
spatial and isospin indices taken separately 
\begin{eqnarray}
\{G^{ia},G^{jb}\} &=& W^{(ij),(ab)} + W^{[ij],[ab]}  \ , 
\label{twoGs}
\end{eqnarray}
where 
\begin{eqnarray}
W^{(ij),(ab)} &=& \frac12 \{G^{ia},G^{jb}\} + \frac12 \{G^{ib},G^{ja}\} \ , \\
W^{[ij],[ab]} &=& \frac12 \{G^{ia},G^{jb}\} - \frac12 \{G^{ib},G^{ja}\} = \epsilon^{ijk}
\epsilon^{abc} A^{kc} \ , 
\end{eqnarray}
with 
\begin{eqnarray}
A^{kc} &=& \frac14 \epsilon^{kij} \epsilon^{abc}  \{G^{ia},  G^{jb}\} \ .
\end{eqnarray}
Now we can use the following identity, see Ref.~\cite{Dashen:1994qi}, to reduce the number of $G$'s by one
\begin{eqnarray} \label{epsG2}
\epsilon^{ijk} \epsilon^{abc}  \{G^{ia},  G^{jb}\}= - (N_c+2) G^{kc} + \frac12 \{
S^k,I^c\} \ .
\end{eqnarray} 
The $W^{(ij),(ab)}$ tensor has $(I,S)=(0,0),(2,0),(0,2),(2,2)$ components. Only the $(0,0)$
component contributes in the nucleon subspace. It is obtained by contracting the indices
and is $W^{ij,ab}_{(0,0)}= \frac{2}{9} \delta^{ij} \delta^{ab} G^{kc}G^{kc} $. The $SU(4)$
quadratic Casimir operator $C_2 = \frac12 S^2 + \frac12 I^2 + 2 G^{kc}G^{kc}$ evaluated in
the symmetric irrep $S_N$ that corresponds to ground state nucleons gives $C_2(S_{N_c}) =
\frac{3}{8} N_c(N_c+4) \mathds{1}$ and shows explicitly that to leading order
$G^{kc}G^{kc}$ can be replaced by the unit operator. The two terms in Eq.~(\ref{epsG2})
that enter in $W^{[ij],[ab]}$ correspond to the $(1,1)$ component, and only the first one
contributes to leading order.

We obtain
\begin{eqnarray}
\langle {\rm N} |  N_c^{-2} G^{ia} G^{jb} |{\rm N} \rangle &=& \langle {\rm N} | \frac{N_c+4}{48 N_c} \delta^{ij} \delta^{ab} \mathds{1} 
-  \frac{N_c+2}{8 N_c} \epsilon^{ijk} \epsilon^{abc} \left( \frac{G^{kc}}{N_c} \right)
   |{\rm N} \rangle + {\cal O}(N_c^{-2}) \ , 
\label{twoGred}
\end{eqnarray}
where this is an example of Eq.~(\ref{eq:manyG}) that shows explicitly the tensor structure in the
spatial and isospin indices.  

\section{Useful tensor identities}

\label{ap-reduction}

Here we collect a few identities involving 
epsilon tensors that are used to simplify the number of spin-flavor and also momentum 
structures. The product of two epsilon tensors can be written as
\begin{eqnarray} \label{eq:epseps}
\epsilon_{ijk} \epsilon_{lmn} &=& {\rm det}
\left[
\begin{array}{ccc}
\delta_{il} & \delta_{im} & \delta_{in} \\
\delta_{jl} & \delta_{jm} & \delta_{jn} \\
\delta_{kl} & \delta_{km} & \delta_{kn} \\
\end{array} 
\right] \nonumber \\
&=& \delta_{il} ( \delta_{jm} \delta_{kn} - \delta_{jn} \delta_{km} ) 
  - \delta_{im} ( \delta_{jl} \delta_{kn} - \delta_{jn} \delta_{kl} ) 
  + \delta_{in} ( \delta_{jl} \delta_{km} - \delta_{jm} \delta_{kl} ) \ .
  \label{eq:doubleepsilon}
\end{eqnarray}
This expression is very useful for constructing an independent set of momentum structures. Another useful identity can 
be obtained by
contracting Eq.~(\ref{eq:doubleepsilon}) with $\epsilon_{ijp} $ giving
\begin{eqnarray}
\delta_{kp} \epsilon_{lmn} &=& \delta_{kn} \epsilon_{lmp} + \delta_{km} \epsilon_{lpn} +
\delta_{kl} \epsilon_{pmn} \ , 
\end{eqnarray}
from where 
\begin{eqnarray}
(A \times B)^i C^j + (B \times C)^i A^j + (C \times A)^i B^j = (A \times B) \cdot C \
\delta^{ij}  
\label{eq:triple}
\end{eqnarray} 
is obtained.
This can be used to eliminate triple products from all our momentum structures and also to reduce the number of 
momentum or spin-flavor structures that contain a cross product.

\end{document}